\documentclass[twocolumn]{aastex6}
\usepackage{natbib}
\usepackage{color}

\def    \apjl  		{\rm {ApJL}}
\def    \apj  		{\rm {ApJ}}
\def    \mnras  	{\rm {MNRAS}}
\def    \araa  		{\rm {ARA\& A}}
\def    \apjl  		{\rm {ApJL}}

\def	\cm		{\,{\rm {cm}}}
\def	\K		{\,{\rm K}}
\def	\g		{\,{\rm {g}}}
\def	\mum	{\,{\mu \rm{m}}}

\def \bea {\begin{eqnarray}}
\def \ena {\end{eqnarray}}

\def	\ted	{{\tau_{\rm ed}}}
\def	\tH	{{\tau_{\rm H}}}

\def    \bmu    {{\hbox{\boldsym\char'026}}}	

\def    \bomega {{\hbox{\boldsym\char'041}}}	

\def	\cm	{\,{\rm cm}}
\def	\d	{{\rm d}}

\def	\D	{{\rm D}}

\def	\ed	{{\rm ed}}

\def	\erg	{\,{\rm erg}}

\def	\g	{\,{\rm g}}

\def	\H	{{\rm H}}

\def	\Hz	{\,{\rm Hz}}

\def	\K	{{\rm K}}

\def	\s	{\,{\rm s}}

\font\mib=cmmib10

\def\bomega{\hbox{\mib\char"21}}

\def\bmu{\hbox{\mib\char"16}}

\begin{document}
\shorttitle{Spinning dust from AGB envelopes}
\shortauthors{Tram, Hoang, Soam, Lesaffre \& Reach}
\title{Modeling rotational disruption of grains and microwave emission from spinning dust in AGB envelopes}

\author{Le Ngoc Tram\altaffilmark{1,2}, Thiem Hoang\altaffilmark{3,}\altaffilmark{4}, Archana Soam\altaffilmark{1}, Pierre Lesaffre\altaffilmark{5,6}, William T. Reach\altaffilmark{1}}

\affil{$^1$ Stratospheric Observatory for Infrared Astronomy, Universities Space Research Association, NASA Ames Research Center, MS 232-11, Moffett Field, 94035 CA, USA; \href{mailto:ngoctram.le@nasa.gov}{ngoctram.le@nasa.gov}}
\affil{$^2$ University of Science and Technology of Hanoi, VAST, 18 Hoang Quoc Viet, Vietnam}
\affil{$^3$ Korea Astronomy and Space Science Institute, Daejeon 34055, South Korea}
\affil{$^4$ Korea University of Science and Technology, 217 Gajeong-ro, Yuseong-gu, Daejeon, 34113, South Korea}
\affil{$^5$ Laboratoire de Physique de l'\'Ecole normal sup\'erieur, ENS, Universit\'e PSL, CNRS, Sorbonne Universit\'e, Universit\'e de Paris, France}
\affil{$^6$ Observatoire de Paris, PSL University, Sorbonne Universit\'e, LERMA, 75014, Paris, France}

\begin{abstract}
Radio observations of some Asymptotic Giant Branch (AGB) star envelopes show the excess emission at frequencies below 100 GHz which cannot be explained by thermal dust emission (hereafter anomalous microwave emission-AME). Moreover, AGB envelopes are a common place where gas molecules condense to form nanoparticles (e.g., polycyclic aromatic hydrocarbons) and large grains. In this paper, we will study whether electric dipole emission from rapidly spinning nanoparticles can reproduce the AME observed toward AGB stars. To properly model the size distribution of nanoparticles in the AGB envelope, we take into account both the increase of nanoparticles due to rotational disruption of large grains spun-up by radiative torques and the decrease of smallest nanoparticles due to rotational disruption driven by stochastic gas collisions. We then perform detailed modeling of microwave emission from rapidly spinning nanoparticles from both C-rich and O-rich AGB envelopes using the grain size distribution constrained by rotational disruption. We find that spinning dust emission is dominant over thermal dust emission at frequencies below 100 GHz. We attempt to fit the observational data of AME using our spinning dust model and demonstrate that spinning dust can reproduce the observed AME in six AGB stars. Finally, we discuss that microwave emission from spinning dust in AGB envelopes could be observed with high-resolution upcoming radio telescopes such as ngVLA and ALMA Band 1. This would be a major leap for understanding AGB envelopes, formation, evolution, and internal structures of dust. Observations would help to distinguish the carrier of AME from comparing C-rich to O-rich stars, because PAHs are formed in C-rich AGB stars while silicates are formed in O-rich stars. 

\end{abstract}

\section{Introduction\label{sec:intro}}
Late in their evolution, low- and intermediate-mass stars (1--8 M$_{\odot}$) reach the Asymptotic Giant Branch (AGB) phase, before they become white dwarfs. 
During this phase, AGB stars lose most of its material \citep{1999isw..book.....L} because radiation pressure accelerates them to speeds above the star’s escape velocity. Mass-loss builds an expanding circumstellar envelope (CSE) around the star, containing dust and gas (e.g., \citealt{Olofsson_2010}; \citealt{Ramstedt_2011}; \citealt{Cox_2012}). 
CSEs of AGB stars can be considered as the most significant chemical laboratories in the universe because their effective temperatures are usually low ($T_{\star}\simeq$ 2000 K - 3500 K), and the mass-loss timescale is long, so that molecules can form in the envelope through chemical and physical processes (e.g., \citealt{Cernicharo_2000}; \citealt{Tenenbaum_2010}).

It is well-known that in the AGB envelopes, simple molecules condense to form complex molecules and then tiny nanoparticles including polycyclic aromatic hydrocarbons (PAHs), and finally to submicron-sized grains before being expelled into the ISM due to radiation pressure. Theoretical studies on the formation of PAHs in the evolved star envelope is well studied (see \citealt{2011EAS....46..177C} and references therein). While observation evidence of dust grains from AGB envelopes is well established thanks to infrared emission from dust grains. However, observational evidence for the existence of PAHs which are demonstrated through mid-IR emission features at 3.3, 6.2, 7.7, 8.6, 11.3, and 17 $\mu$m (\citealt{1984A&A...137L...5L}; \citealt{1985ApJ...290L..25A}; \citealt{2007ApJ...656..770S}; \citealt{2007ApJ...657..810D}) is not yet available (see \citealt{2008ARA&A..46..289T} for a review). The underlying reason for that is during the AGB stage, the star temperature is rather low, such that there lacks of UV photons to trigger mid-IR emission of PAHs. One note that prominent PAH features are usually observed at the later stage such as planetary nebula due to higher temperatures of the central star. Therefore, in order to achieve a complete understanding on the formation of dust in AGB star envelopes, it is necessary to seek a new way to observe PAHs/nanoparticles in these environments.

Modern astrophysics establishes that rapidly spinning nanoparticles that have permanent electric dipole moment (e.g., PAHs, nanosilicates, nanoiron particles) can emit electric dipole radiation at microwave frequencies below 100 GHz (\citealt{1998ApJ...508..157D}; \citealt{Hoang:2010jy}; \citealt{2016ApJ...824...18H}; \citealt{2016ApJ...821...91H}). The rotational excitation of nanoparticles can be achieved in the absence of UV photons, such that microwave emission from spinning dust can be efficient in the environments without UV photons like AGB envelopes and dense shocked regions (\citealt{2019ApJ...877...36H}; \citealt{2019ApJ...886...44T}). Therefore, microwave emission could be a unique way for us to probe nanoparticles in the envelope of AGB stars. In particular, with upcoming advanced radio telescopes, e.g., ALMA Band 1, the Square Kilometer Array (SKA), and next generation VLA (ngVLA), the potential of observing nanoparticles via spinning dust emission is very promising, which would allow us to have a better understanding on their formation and evolution in general. 

Interestingly, previous radio observations show the anomalous microwave emission (AME) at frequencies below 100 GHz from the CSEs around some AGB stars that cannot be explained by thermal dust emission, such as from IRC +10216 (\citealt{1989A&A...220...92S} and \citealt{2006A&A...453..301M}), or from VY CMa, IRc +10216, CIT 6 and R Leo (\citealt{1995ApJ...455..293K}), or from $\alpha$ Her, IRC +10216, IRC +20370, WX Psc, $\alpha$ Sco, V Hya (\citealt{2007MNRAS.377..931D}). To date, the origin of such excess is not well understood. Therefore, the goal of this paper is to perform detailed modeling of spinning dust emission from nanoparticles in AGB stars and explore whether spinning dust mechanism can explain such AME. 


To model the emission from spinning dust, we consider the radiation-driven wind model, which starts from dust-formation zone (i.e., $\sim$ 5-10R$_{\ast}$), where the temperature varying from 1000 K down to 600 K and a total number density varying from $10^{10}$ to $10^{8}$ cm$^{-3}$. Furthermore, dust grains are classified by their spectral features and they correspond to a special kind of envelope properties. Silicates, identified at 9.7 and 18 $\mu$m are considered as a major feature of oxygen-rich stars ($C/O<1$) (\citealt{Kwok_2004}). In contrast, in carbon-rich stars ($C/O>1$), one expect to observe prominent mid-IR features from PAHs. In this paper, we will model the microwave emission from spinning PAHs for a typical representative of carbon-rich AGB stars (IRC +10216) and from nanosilicates for a typical representative of oxygen-rich AGB stars (IK Tau). 

Owing to their large sizes, AGB stars are among the most luminous objects in the sky, with luminosities of order $10^{4} L_{\odot}$. Thus, the radiation field in their CSEs is remarkable strong. Recently, \cite{2019NatAs...3..766H} showed that a strong radiation field can torque large dust grains 
up to $\sim 10^{9}$ revolutions per second, at which rate the centrifugal force can disrupt a large grain into smaller ones. In addition, \cite{2019ApJ...877...36H} and \cite{2019ApJ...886...44T} showed that very small grains (referred to nanoparticles whose sizes are lower 10 nm) can be spun up to suprathermal rotation by stochastic torques induced by gas bombardment in a dense and hot medium. As a result, smallest nanoparticles are disrupted into smaller tiny fragments when the centrifugal force excesses the material strength, after which dust particles are thermally sublimated into the gas phase (\citealt{1989ApJ...345..230G}). These two disruption mechanisms are then taken into account in our model because they can modify the size distribution of dust grains in AGB envelopes.

The structure of this paper is as follows. Section \ref{sec:AGB_model} describes the parametric physical model of AGB envelopes. We constrain the upper limit of size of grains by RATD mechanism in Section \ref{sec:RATD}, and the lower limit of size of nanoparticles by mechanical torques in Section \ref{sec:mechanical_torques}. In Section \ref{sec:spin_model}, we review the spinning dust model, and calculate the SED from spinning dust for two examples of AGB stars. An extended discussion on implications of our results for explaining AME in AGB envelopes in Section \ref{sec:discuss}. A summary of our main results is presented in Section \ref{sec:sum}.

\section{Physical model of AGB envelopes} \label{sec:AGB_model}

During the AGB phase, a star loses most of its mass through stellar winds. Very close to the central star, where it is too hot for dust grains to form, the ejection mechanism is not well unified yet. The pulsation process is believed to accelerate gas above the escape velocity, sending it radially outward (e.g., \citealt{1997A&A...319..648H}, \citealt{2000upse.conf..227W}; and see \citealt{2018arXiv180801439T} for review). Farther from the star, where the gas temperature decreases to the condensation temperatures, dust grain form. Subsequently, dust grains are accelerated by radiation pressure and transfer their momentum to gas molecules through collisions. Collisions result in a drag force \citep{1972ApJ...178..423G} that allows gas beyond the dust-formation distance to overcome the gravitational well of the star. The resulting winds are called dust-driven or radiation-driven winds. Here we briefly describe our parametric physical model of AGB envelopes for a radiation-driven wind, which will be used as the 
framework for calculating rotational disruption of dust grains and spinning dust emission.

\subsection{Gas density profile}
Let $\dot{M}$ be the rate of mass loss of an AGB star, and we assume a spherical envelope. 
The gas density at distance $r$ is then given by
\begin{equation}
\begin{array}{r@{}l}
n_{\rm gas}(r) &{}= \frac{\dot{M}}{4\pi r^{2}m_{\rm gas}v_{\rm exp}} \\
 &{}=   10^{6} \left(\frac{\dot{M}}{10^{-5}M_{\odot}}\right) \left(\frac{10^{15}{\rm cm}}{r}\right)^{2}
\left(\frac{10\,{\rm  km~s}^{-1}}{v_{\rm exp}}\right)\,{\rm cm}^{-3}
\end{array}
\end{equation}
where $v_{\rm exp}$ is the expansion velocity of the outflow. The typical mass loss rates are $\dot{M}\sim 1$--$10 \times 10^{-5}$ $M_{\odot}$~yr$^{-1}$. 
For cold AGB stars, we take $m_{\rm gas}=2.33$ amu because hydrogen in the CSE around these stars is predominantly molecular (\citealt{1983MNRAS.203..517G}; \citealt{2018arXiv180801439T}). For hot AGB stars, on the contrary,  the atomic form of hydrogen  dominate the CSE, therefore we take $m_{\rm gas}=1.3$ amu (90\% H and 10\% He). The critical temperature to distinguish cold or hot AGB stars depends on the density of stellar photosphere, e.g., $T^{\rm cri}_{\ast}\simeq 2500\K$ for stars with density $\geq 10^{14}\cm^{-3}$ (\citealt{2018arXiv180801439T}). The expansion velocity is likely constant for a radiation-driven wind model as demonstrated by  \cite{1983ApJ...271..702T} and \cite{1994A&A...290..573K}.


\subsection{Gas Temperatures}
In the envelope beginning at the condensation radius ($r_{c}$), where the gas expands adiabatically at constant $v_{\exp}$, the radiative heating is dominant because the gas is weakly or not longer shielded by dust so that the temperature drops gradually with the distance, and we approximate its profile as a power law as: (e.g., \citealt{1988ApJ...328..797M}; \citealt{2006ApJ...650..374A}; \citealt{2010A&A...516A..69D}; \citealt{2016A&A...588A...4L}):
\bea
T_{\rm gas}(r)= T_0\left(\frac{r}{r_0}\right)^{-\alpha_{\rm gas}},
\ena
where $T_0$, $r_0$, and $\alpha_{\rm gas}$ are parameters whose values are given in Table \ref{tab:wind_params}.

\subsection{Grain temperature}
The grain temperature evolution is determined by the balance between the heating and  cooling rates. Grains are heated  by collisions with the gas particles and by absorption of stellar or ambient radiation. Grains are cooled by collisional energy transfer and by thermal radiation. The detailed profile of dust temperature in a realistic envelope can be modeled by solving the radiative transfer equation for a given dust opacity (\citealt{Winters_1994}). But we assume here that the stellar radiation dominates. In the simple case when the grain absorption efficiency can be approximated as a power-law of waveleght $Q_{\rm abs} \sim \lambda^{-s}$, the dust temperature can be derived as (\citealt{2003agbs.conf.....H}): 
\bea \label{eq:Td}
T_{\d}(r)=T_{\ast}\left(\frac{R_{\ast}}{2r}\right)^{2/(4+s)},
\ena
where $T_{\ast}$ and $R_{\ast}$ are the temperature and radius of star. Observations suggest $s\simeq 1$. 

\subsection{Physical properties of a C-star and an O-star}
\begin{figure}
\includegraphics[width=0.5\textwidth]{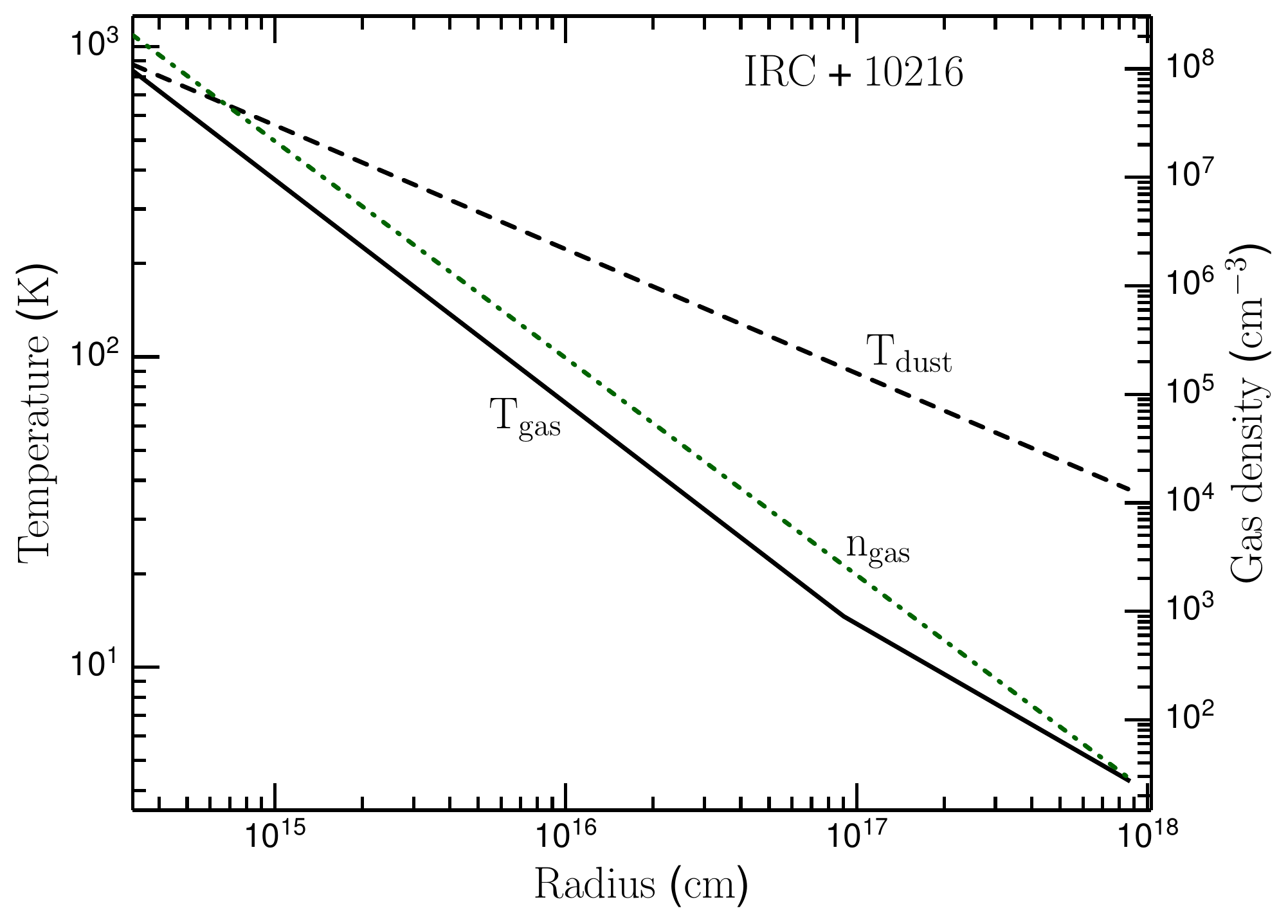}
\includegraphics[width=0.5\textwidth]{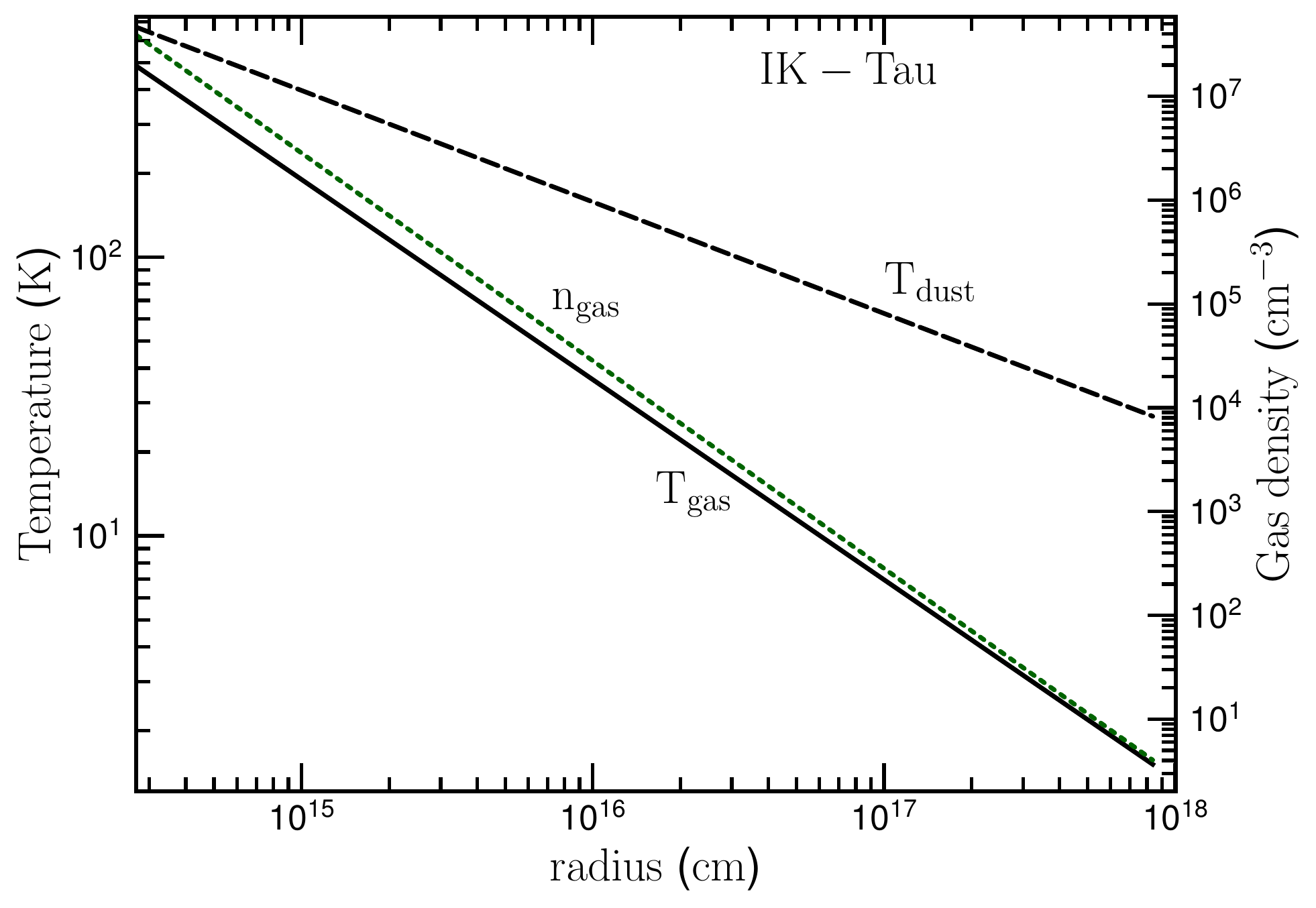}
\caption{Physical properties of freely expansion wind in CSE around IRC +10216 star (top pannel) and IK Tau (bottom panel).}
\label{fig:phys_free}
\end{figure}

We apply the parametric models described above to calculate the physical properties of the freely expanding wind in CSE around the well-known C-rich AGB stars (namely IRC +10216) and O-rich (namely IK Tau) as illustrated in Figure \ref{fig:phys_free}. For IRC +10216, the profile of gas temperature is taken from \cite{1988ApJ...328..797M}, who fitted the power-law to the theoretical results of \cite{1982ApJ...254..587K}. For IK Tau, we similarly fit the power-law to the gas temperature profile that is derived from observational $^{12} \rm CO$ lines (\citealt{2010A&A...516A..69D}). Other parameters are listed in Table \ref{tab:wind_params} with the references given.    

\begin{table}
\centering
\caption{Circumstellar envelope models}\label{tab:wind_params}
\begin{tabular}{lll} 
\hline\hline
Parameters & IRC +10216 & IK Tau \\[1mm]
\hline
$L_{\ast}$ ($L_{\odot}$) & $^{(1)}$8640 & $^{(9)}$7700\\
$\dot{M}$ ($M_{\odot}$~ yr$^{-1}$) & $^{(2)}$2$\times 10^{-5}$ & $^{(10)}$ $4.5\times 10^{-6}$ \\
$T_{\ast}$\ (K) & $^{(3)}$ 2200 & $^{(9,10)}$2100\\ 
$T_0$\ (K) & $^{(4)}$14.6 & $^{(11)}$707 \\
$R_{\ast}$\ (cm) & $^{(5)}$$4.5\times 10^{13}$ & $^{(10)}$$3.1\times 10^{13}$\\
$r_{c}$\ ($R_{\ast}$) & $^{(6)}$5 & ${(12)}$8.7\\
$r_0$\ ($R_{\ast}$) & $^{(4)}$1385 & $^{(11)}$5.20\\
$r_{out}$\ ($R_{\ast}$) & $^{(7)}$$1.32\times 10^{4}$ & $^{(12)}$$2\times 10^{4}$\\
$v_{exp}$\ (km~s$^{-1}$) & $^{(8)}$14.5 & $^{(13)}$24\\
$\alpha_{\rm gas}$ & $^{(4)}$0.72 ($r < r_{0}$) & $^{(11)}$0.79\\
               & $^{(4)}0.54$ ($r\geq r_{0}$) & \\
$D$ (pc) & $^{(5)}$130 & $^{(9)}$265 \\
\hline\hline\\
\end{tabular}
\noindent
    {\bf \\ References.} (1) \cite{2012A&A...543A..73M}; (2) \cite{1997ApJ...483..913C}; (3) \cite{2015MNRAS.449..220M}, (4) taken from the fit of \cite{1988ApJ...328..797M} to \cite{1982ApJ...254..587K}; (5) \cite{2018arXiv180801439T}; (6) \cite{2006ApJ...650..374A}; (7) \cite{2010ApJ...711L..53S}; (8) \cite{1998ApJS..117..209K}; (9) \cite{2016A&A...591A..44M}, (10) \cite{2010A&A...523A..18D} , (11) fit to \cite{2010A&A...516A..69D}, (12) \cite{2010A&A...516A..69D}, (13) \cite{2012A&A...537A.144J}
\end{table}

\subsection{Drift velocity for nanoparticles}
Once dust grains have formed, they scatter and absorb stellar photons, leading to a radiative force that pushes them outwards away from the star. The radiation force on a grain of radius $a$ at distance $r$ due to the central star is given by
\bea
F_{\rm rad}=\frac{L_{\star}\bar{Q}_{\rm rp}\pi a^{2}}{4\pi r^{2}c}.\label{eq:Frad}
\ena
where $c$ is the speed of light, $Q_{\rm rp}$ is the radiation pressure efficiency, and $\bar{Q}_{\rm rp}$ is the wavelength averaged radiation pressure efficiency weighted by the stellar spectrum, and $L_{\star}$ is the total stellar luminosity. Let $v_{\rm drift}$ be the drift velocity of grains through the gas. If the drift speed is much larger than the thermal speed of gas particles, the drag force per unit volume of the gas is $F_{\rm drag}(a) = \pi a^{2}\rho v_{\rm drift}^2 n_{\rm gas}$, where a is the radius of a single grain.\footnote{When the drift velocity is above a percent of the light speed, the drag force decreases with increasing the velocity, as discovered in \cite{2017ApJ...847...77H}.} On the other hand, if the drift speed is lower than the thermal speed of the gas, the drag force $F_{\rm drag}(a) = \pi a^2 \rho c_s v_{\rm drift} n_{\rm gas}$. To combine those limits, we can express the drag force as follows:
\bea
F_{\rm drag}(a) = \pi a^2 n_{\rm gas} \rho v_{\rm drift} \sqrt{v_{\rm drift}^{2}+c^{2}_{s}}\rm{.}
\ena
Because the mean free path of the gas is higher than the typical dust radii, and the velocities of gas and dust are different, the grains are not position coupled to the gas. Despite the fact that the grains collide with only a small fraction of the gas particles, \cite{1972ApJ...178..423G} indicates that the subsequent collisions among the gas molecules allow the momentum that they receive from the radiation field to be transferred to the gas. \cite{1972ApJ...178..423G} also demonstrates that the small grains rapidly reach the terminal drift velocity. The grains move at the terminal drift velocity when the radiation force balances the drag force:
\bea
    v^{2}_{\rm drift}=\frac{1}{2}\left[\left[ \left( \frac{2v}{\dot{M}c_l}\bar{Q}_{\rm rp}L_{\star}\right)^{2} + c^{4}_{s}\right]^{0.5} -c^2_{s}\right]
\ena
where $c_{s}=(2k_{B}T_{\rm gas}/m_{\rm gas})^{1/2}$ is the sound speed in the gas. We define the dimensionless drifting parameter $s_{d}=v_{\rm drift}/c_{s}$. As a limitation, if the thermal speed is much smaller than speed of outflow and the grain size is much smaller than wavelength, $\bar{Q}_{rp} \propto a$ and therefore $s_{d}(r) \sim v_{\rm drift}(r) \propto a^{0.5}$, which means the drift between gas and big dust grains is larger than that of small dust grains.

\begin{figure}
\includegraphics[width=0.45\textwidth]{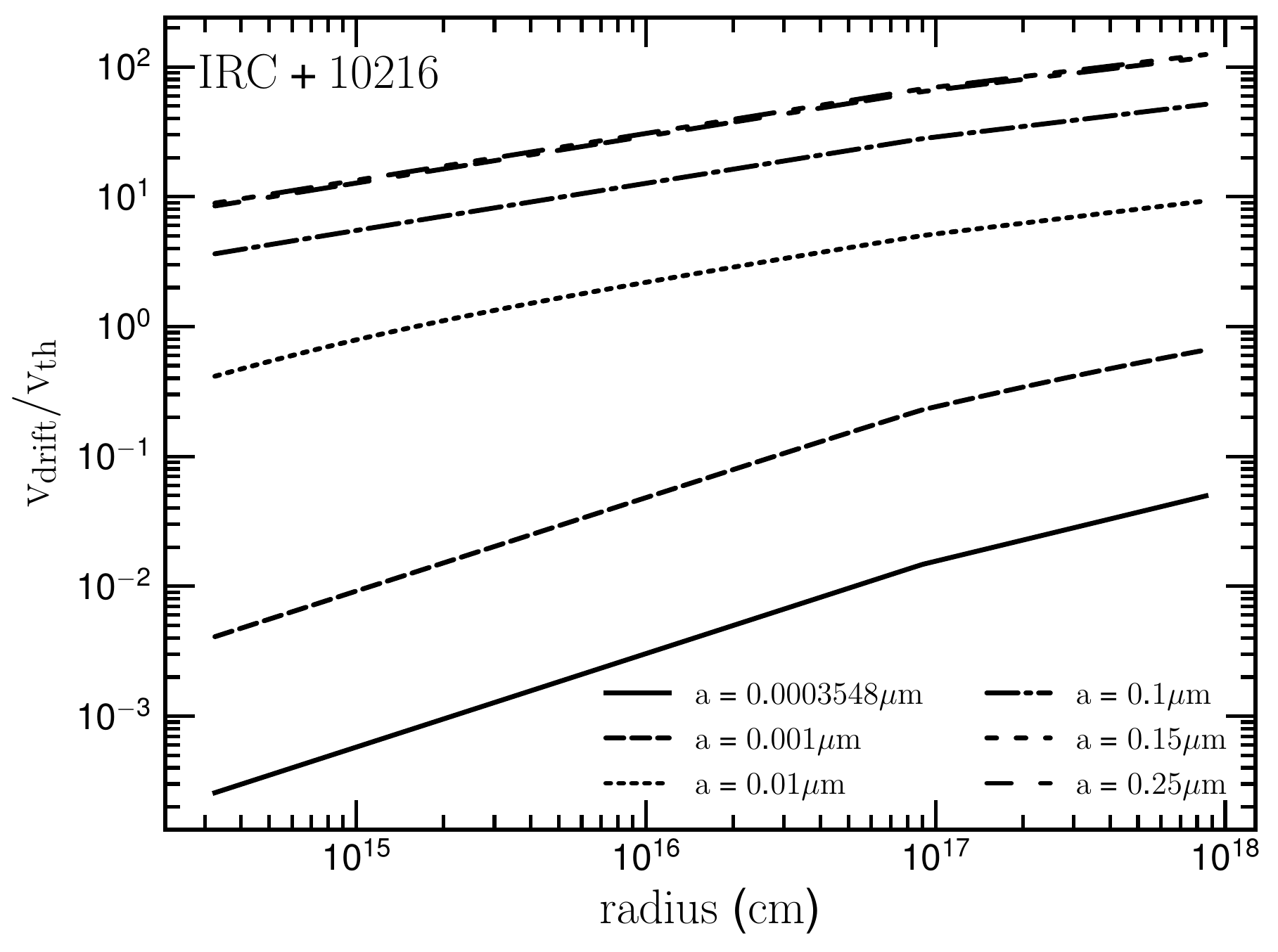}
\includegraphics[width=0.45\textwidth]{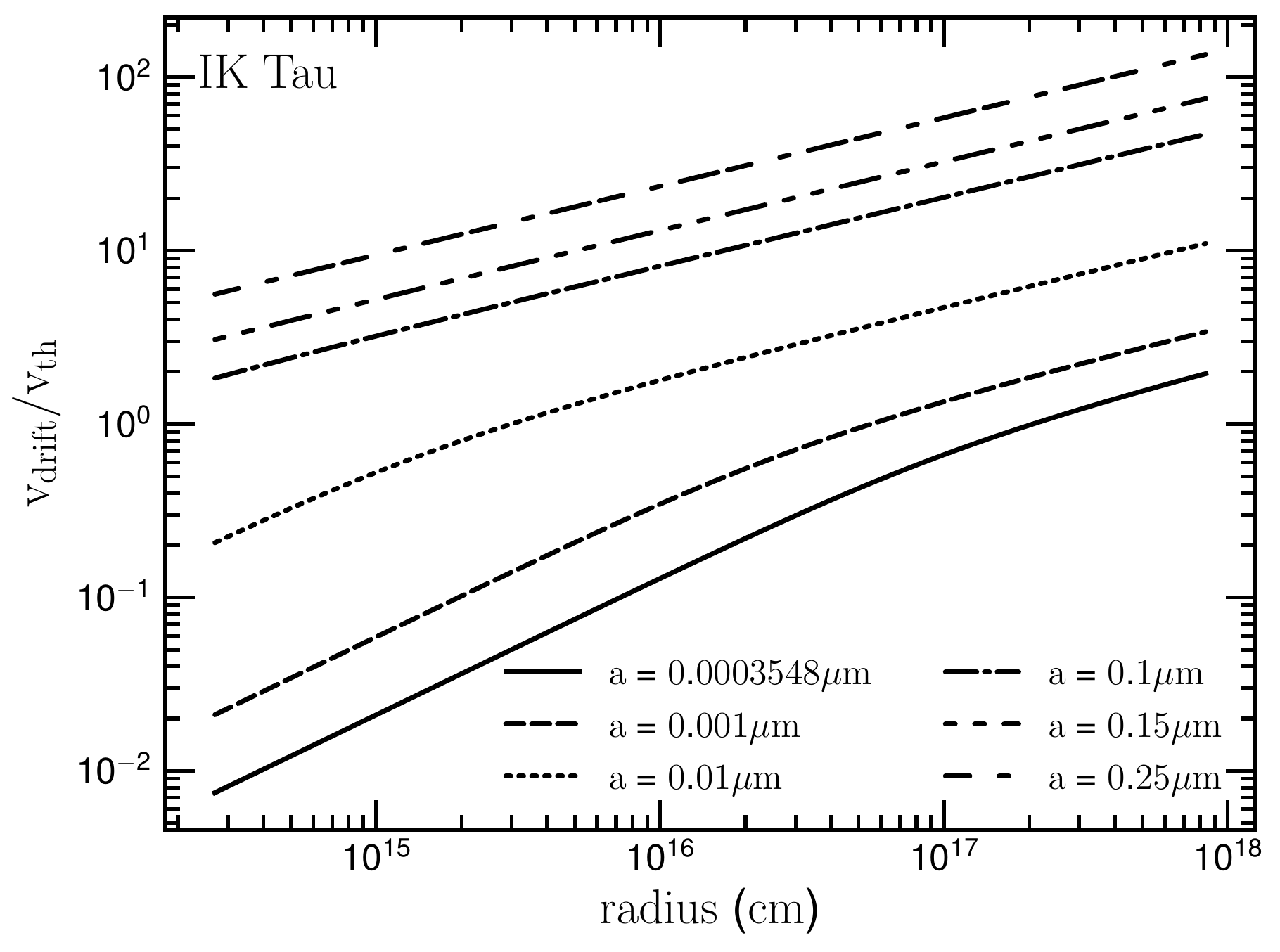}
\caption{Dimensionless drifting parameter for IRC +10216 star (top panel) and IK Tau (bottom panel) calculated for the different grain sizes. Large grains are moving supersonically, but smallest nanoparticles are moving subthermally.}
\label{fig:sd_free}
\end{figure}

In Figure \ref{fig:sd_free}, we show the normalized grain drift velocity $v_{\rm drift}/v_{\rm th}$ calculated for the different grain sizes. The drift velocity $s_{d}$ tends to increase with increasing the distance because of the decrease of the gas temperature. Large grains can be accelerated to supersonic motion, while smallest nanoparticles of size $a\lesssim 1$ nm are mostly moving at subsonic speeds.

\subsection{Ionization}
In the AGB and post-AGB stages, the star is relatively cool, with $T_{\star}\sim 2000-3000$ K. Therefore, its ionizing radiation is negligible, and the gas remains
mostly neutral. In the stage of PN, the star has contracted to a white
dwarf with high temperature, and the envelope is ionized. Nanoparticles in the outer envelope will be affected by interstellar UV photons which also ionize gas atoms. The most abundant species in C-rich and O-rich stars are listed in Table \ref{tab:most_abundant} with references.

\begin{table*}
\centering
\caption{Abundance relative to $\rm H_{2}$ of major chemical species in AGB envelope.}\label{tab:chem_component}
\begin{tabular}{ll}
\begin{tabular}{lll}
 & IRC +10216 &\\
\cline{1-3}
\cline{1-3}
Species & abundances &Note\\
\cline{1-3}
x(H$_{2}$) & $^{(1)}$ $\approx 1$ &    \\
x(He) & $^{(2)}$ 2$\times$ 10$^{-1}$ & \\
x(CO) & $^{(2)}$ 6$\times$ 10$^{-4}$ &  \\
x(C$^+$) & $^{(3)}$ 1.5$\times$ 10$^{-5}$ & r$<$ 6$\times$ 10$^{17}$\ \cm \\
x(H$^+$) & $^{(3)}$ 3.7$\times$ 10$^{-8}$ & $r\geq$ 6$\times$ 10$^{17}$\ \cm \\
x(S$^+$) & $^{(3)}$ 5$\times$ 10$^{-8}$ & $r\geq$ 3$\times$ 10$^{16}$\ \cm \\
&&\\
\end{tabular}
&
\begin{tabular}{lll}
 & IK Tau &  \\
\cline{1-3}
\cline{1-3}
Species & abundances &Note\\
\cline{1-3}
x(H$_{2}$) & $^{(1)}$ $\approx 1$ &  \\
x(CO) & $^{(4)}$ $\times$ 10$^{-4}$ & r$<$ 2.5$\times$ 10$^{17}$\ \cm\\
x(O) & $^{(4)}$ 1$\times$ 10$^{-4}$ & $r\geq$ 10$^{16}$\ \cm\\
x(N) & $^{(4)}$ 2.5$\times$ 10$^{-4}$& $r\geq$ 2$\times$ 10$^{17}$\ \cm \\
x(N$_{2}$) & $^{(4)}$ 1.8$\times$ 10$^{-4}$ & $r\leq$ 6$\times$ 10$^{17}$\ \cm \\
x(C$^+$) & $^{(4)}$ 2$\times$ 10$^{-4}$ \\
x(e$^-$) & $^{(4)}$ 2$\times$ 10$^{-4}$ \\
\end{tabular}
\end{tabular}
\centering
\\ {\bf References.} (1)\cite{1983MNRAS.203..517G}, \cite{2018arXiv180801439T}; (2) \cite{2006ApJ...650..374A}; (3) \cite{2000MNRAS.316..195M}; (4) \cite{2016A&A...588A...4L}  
\label{tab:most_abundant}
\end{table*}


\section{Rotational disruption of large grains by radiative torques} \label{sec:RATD}
 \subsection{Radiation field from an AGB star}
In this paper, we are considering the radiative-driven wind, which should start from the dust condensation zone. Normally, this zone is located quite far from the central star (e.g., $\simeq$ 5$R_{\ast}$ for the case of IRC +10216). Thus we can neglect the angle-dependence of radiation field to dust grain particles. Therefore, the radiation energy density from an AGB star at a distance $r$ is:
\bea
u_{\rm rad} \simeq 0.46 \frac{L_4 e^{-\tau}}{r^{2}_{au}}\ \rm{(\erg\cm^{-3})}
\ena
where $L_4=L_{\ast}/10^{4}\,L_{\odot}$, and $\tau$ is the total optical depth. 
The averaged interstellar radiation field (ISRF) from other Milky Way stars
has energy density $u_{\rm ISRF}=8.64 \times 10^{-13} \erg \cm^{-3}$ (\citealt{1983A&A...128..212M}).
The relative strength of the radiation field from the AGB star is then
\bea
U\equiv  \frac{u_{\rm rad}}{u_{\rm ISRF}}\simeq 5.3\times 10^{11} \frac{L_4 e^{-\tau}}{r^2_{au}} 
\ena
For IRC +10216 at $5R_{\ast}$, $U\simeq 2 \times 10^{9}$, whereas at $10^{4}R_{\ast}$ is $U\simeq 3\times 10^{2}$ (neglecting extinction).

\subsection{Rotational disruption by radiative torques}
As revealed in \cite{2019NatAs...3..766H}, a strong radiation field can spin large grains up to a maximum angular velocity:
\bea \label{eq:omega_rat1}
\omega_{\rm RAT}=\frac{\Gamma_{\rm RAT}\tau_{\rm damp}}{I}
\ena
where $I$ is the grain inertial moment, $\Gamma_{\rm RAT}$ is the radiative torque (\citealt{1996ApJ...470..551D}; \citealt{2007MNRAS.378..910L}; \citealt{Hoang:2008gb}; \citealt{Herranen:2019kj}), and $\tau_{\rm damp}$ is the total damping timescale. 

The averaged radiative torque applied on a grain of irregular shape with {\it effective} size $a$ is (see \citealt{2019NatAs...3..766H} and reference theirin):
\bea
\Gamma_{\rm RAT}=\pi a^{2} \gamma u_{\rm rad}\frac{\bar{\lambda}}{2\pi}\bar{Q}_{\Gamma}
\ena
where $\gamma$ is the  degree of anisotropy of the radiation field ($0\leq \gamma \leq 1$), $\bar{\lambda}$ is the average wavelength of the radiation field (i.e., $\bar{\lambda} = 2.42\mu$m for IRC +10216 and $\bar{\lambda} = 2.53\mu$m for IK Tau). Above, the radiative torque efficiency averaged over the radiation spectrum is approximately $\bar{Q}_{\Gamma}\simeq 2 (\bar{\lambda}/a)^{-2.7}$ for $a \leq \bar{\lambda}/1.8$, and $\bar{Q}_{\Gamma} \sim 0.4$ for $a > \bar{\lambda}/1.8$. However, we adopt the maximum of grain size to be $0.25\mu$m as deduced from observations (\citealt{1977ApJ...217..425M}). Therefore, we just consider  $a\leq \bar{\lambda}/1.8$ in this work. 

The total rotational damping of dust grains consists of collisional damping due to collisions with gas species (\citealt{1967ApJ...147..943J}) and the IR damping due to IR emission, and the characteristic damping time is described by (see \citealt{2019NatAs...3..766H}):  
\bea
\tau_{\rm damp} = \frac{\tau_{\rm gas}}{1+F_{\rm IR}},
\ena
where the characteristic timescale of collisional damping is
\bea
\tau_{\rm gas} \simeq 8.74\times 10^4 a_{-5} \hat{\rho} \left( \frac{30\ \rm{cm^{-3}}}{n_{\H}}\right) \left( \frac{100\K}{T_{\rm gas}}\right)^{1/2} \rm{yr},
\ena
with $n_{\H}=n(\H)+2n(\H_{2})$ being the proton number density, and the dimensionless IR damping coefficient
\bea
F_{\rm IR} \simeq \left(\frac{0.4U^{2/3}}{a_{-5}}\right) \left(\frac{30\ \rm{cm^{-3}}}{n_\H} \right) \left(\frac{100\K}{T_{\rm gas}} \right)^{1/2}~,
\ena
where $a_{-5} = a/(10^{-5}\rm cm)$, and $\hat{\rho}=\rho/(3\ \g \cm^{-3})$. Here we have assumed that dust grains are in the thermal equilibrium established by radiative heating of starlight and radiative cooling by IR emission.

Plugging $\Gamma_{\rm RAT}$ and $\tau_{\rm damp}$ into Equation (\ref{eq:omega_rat1}), yields the maximum rotation rate spun-up by RATs:
\bea \label{eq:omega_rat2}
\omega_{\rm RAT} &\simeq& 6.6\times 10^9 \gamma a^{0.7}_{-5} \bar{\lambda}^{-1.7}_{2.42} \\ \nonumber
&&\times \left(\frac{U_{6}}{n_5 T^{1/2}_{2}}\right) \left(\frac{1}{1+F_{\rm IR}}\right) ~\rm{rad}\s^{-1},
\ena
with $U_6=U/10^6$, $n_5=n_{\H}/10^{5}\cm^{-3}$, $T_{2}=T_{\rm gas}/100\K$, and $ \bar{\lambda}_{2.42}=\bar{\lambda}/2.42 \mu$m. 

Equation (\ref{eq:omega_rat2}) reveals that the rotational rate induced by RATs depends on the term $U/n_{\H}T^{1/2}_{\rm gas}$ and $F_{\rm IR}$. Thus, grains at $154R_{\ast}$ in CSE of IRC +10216, where $U = 10^{6}$, can be spun-up to $\omega_{\rm RAT} \sim 10^{9} \rm rad\s^{-1}$. However, when a grain of size of $a$ is rotating at an angular velocity $\omega$, it develops a centrifugal stress due to the centrifugal force which scales as $S=\rho a^{2} \omega^{2}/4$ (\citealt{2019NatAs...3..766H}). Then, if the rotation rate increases to a critical limit such that the stress induced by centrifugal force exceeds the tensile strength of the material, grains are disrupted instantaneously. The critical angular velocity for the disruption is given by:
\bea
\frac{\omega_{\rm cri}}{2\pi}&=&\frac{1}{\pi a}\left(\frac{S_{\rm max}}{\rho} \right)^{1/2}\nonumber\\
&\simeq& 1.8\times 10^{9}a_{-5}^{-1}\hat{\rho}^{-1/2}S_{\rm max,10}^{1/2}\Hz,~~~~\label{eq:omega_cri}
\ena
where $S_{\max}$ is the maximum tensile strength of dust material and $S_{\max,10}=S_{\max}/(10^{10}\erg\cm^{-3})$.\footnote{An equivalent unit for the tensile strength is ${\rm dyne/cm}^{2}$; in this paper we use the unit of $\erg\cm^{-3}$ for convenience.} 

The exact value of $S_{\max}$ depends on the dust grain composition and structure. Compact grains can have higher $S_{\max}$ than porous grains. Ideal material without impurity, such as diamond, can have $S_{\max}\ge 10^{11}\erg\cm^{-3}$. \cite{1974ApJ...190....1B} suggested that $S_{\max}\sim 10^{9}-10^{10}\erg\cm^{-3}$ for polycrystalline bulk solids (see also \citealt{1979ApJ...231...77D}). Composite grains as suggested by \cite{1989ApJ...341..808M} would have much lower strength. In \cite{1979ApJ...231...77D}, the value $S_{\max}=5\times 10^{9}\erg\cm^{-3}$ is taken for small graphite grains. In the following, nanoparticles with $S_{\max} \gtrsim 10^{10}\erg\cm^{-3}$ are referred to as strong materials, and those with $S_{\max} < 10^{10}\erg\cm^{-3}$ are called weak materials.

For each location in the AGB envelope with given $n_{\H}, T_{\rm gas}$, and $U$, one can obtain the critical grain size of rotational disruption by radiative torques by setting $\omega_{\rm RAT} \equiv \omega_{\rm cri}$.

\begin{figure}
\includegraphics[width=0.45\textwidth]{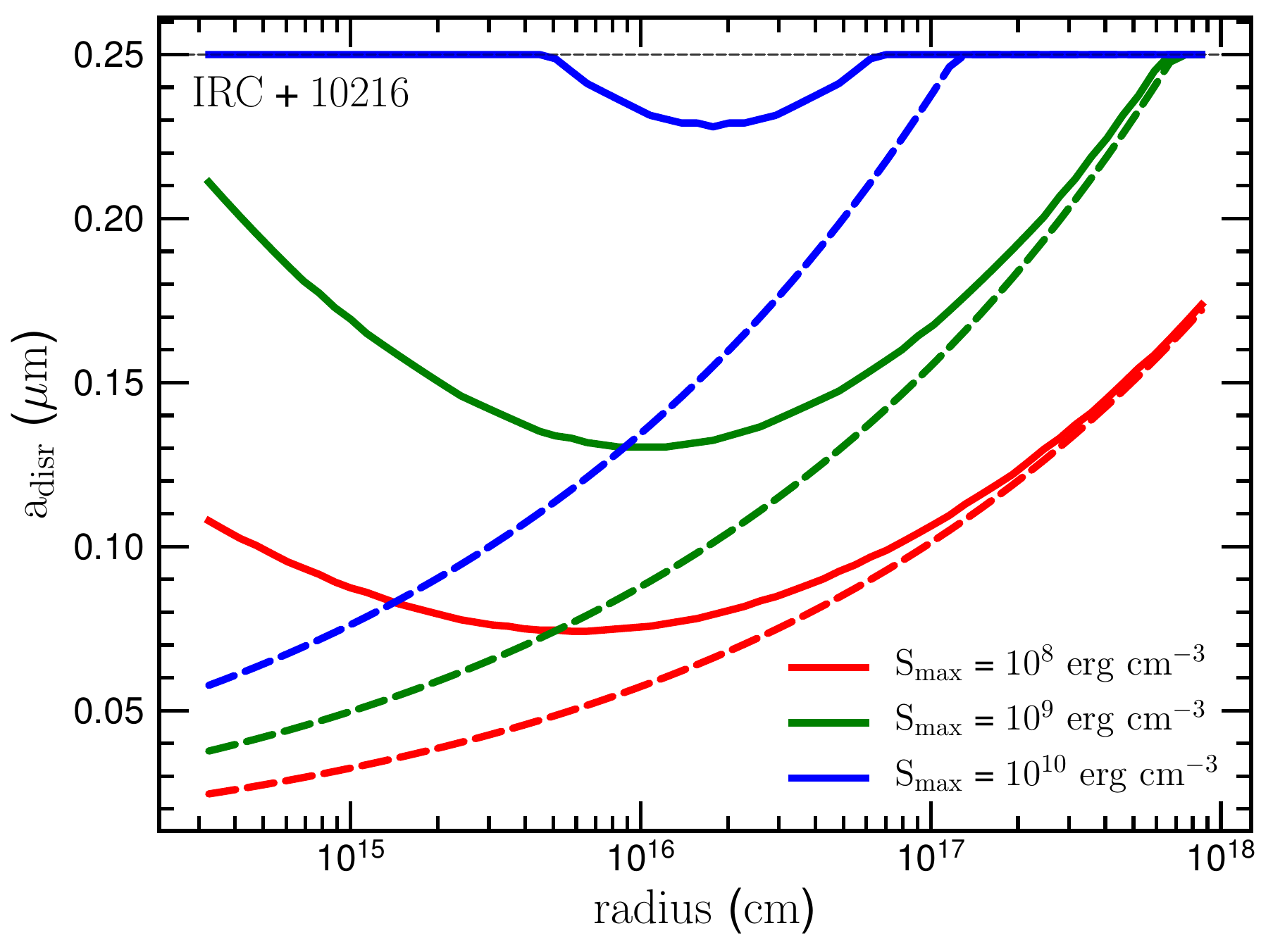}
\includegraphics[width=0.45\textwidth]{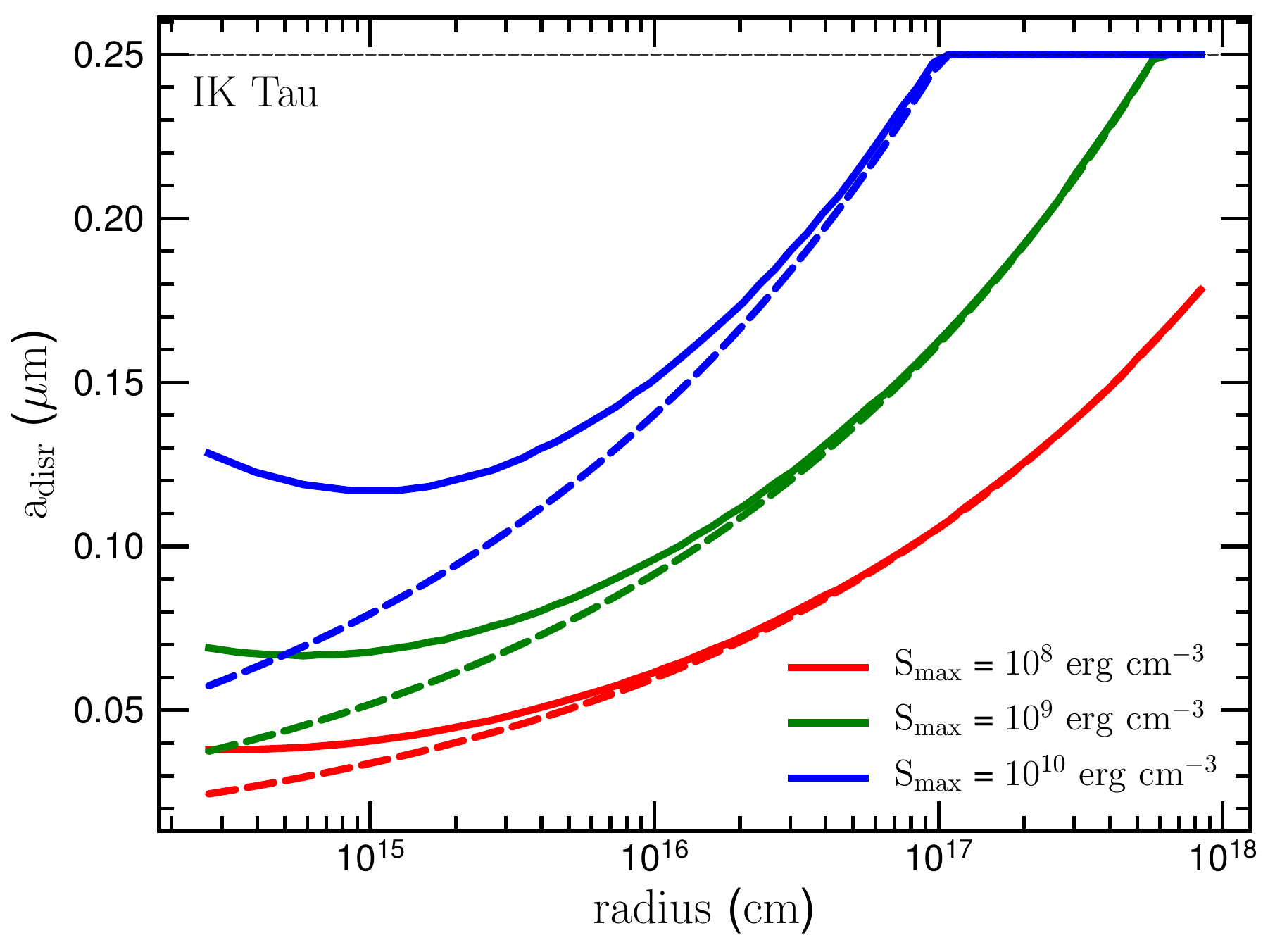}
\caption{Disruption size of dust grains versus distance from the star
in the CSE of IRC +10216 (top panel) and IK Tau (bottom panel) for the different tensile strengths ($S_{\rm max}$). The initial maximum grain size is chosen as $0.25 \mu$m. The corresponding colored dashed lines are disruption size computed by Eq. \ref{eq:adist_analy}. The disruption size is smaller for weaker grain materials (i.e., lower $S_{\rm max}$).}
\label{fig:RATD}
\end{figure}

Figure \ref{fig:RATD} shows the disruption size of dust grains by radiative torques with respect to distance from C-rich star IRC +10216 and O-rich star IK Tau, assuming the different tensile strengths. 
Very close to the star, rotational damping by gas collisions is very efficient because $n_\H \gg 1$ and $T_{\rm gas} \gg 1$ and dominates over IR damping (i.e., $F_{\rm IR} \ll 1$), resulting in a very short damping timescale ($\tau_{\rm damp} \sim \tau_{\rm gas}\ll 1$ yr). As a result, grains lose their angular momentum rapidly and cannot be spun up to critical rotation rates, and large grains can survive rotational disruption. Farther away from the star, the rotational damping rate is decreased due to the rapid decrease of $n_{\H}$ and $T_{\rm gas}$ with distance $r$, large grains can be spun-up to extremely fast rotation by radiative torques and are disrupted into small sub-fragments. The disruption size decreases with increasing distance from the star. Finally, when grains are far enough from the star, the collisional damping is subdominant compared to IR damping, i.e., $F_{\rm IR} \gg 1$ and $U \gg 1$ and the disruption size of grains can be analytically estimated as (\citealt{2019NatAs...3..766H}; \citealt{2019ApJ...876...13H}):
\bea \label{eq:adist_analy}
\left( \frac{a_{\rm disr}}{0.1\ \rm{\mu m}} \right)^{2.7} \simeq 2.12 \gamma^{-1} \bar{\lambda}_{2.42}^{1.7} \left( \frac{U_6}{1.2}\right)^{-1/3} S_{\max,10}^{1/2},
\ena
which does not depend on the gas properties as shown by the dashed lines. 

\section{Rotational disruption of nanoparticles by stochastic mechanical torques} \label{sec:mechanical_torques}
\subsection{Rotational dynamics of nanoparticles}
In dense regions, such as the CSE of AGB stars, \cite{1967ApJ...147..943J} showed that collisions with gas atoms and molecules rotate dust grains. In addition, nanoparticles are  directly bombarded by molecules and atoms (including elements heavier than H), and they experience long-distance interactions with passing ions (\citealt{1998ApJ...508..157D}; \citealt{Hoang:2010jy}; \citealt{2019ApJ...877...36H}). 

\begin{figure}
\includegraphics[width=0.45\textwidth]{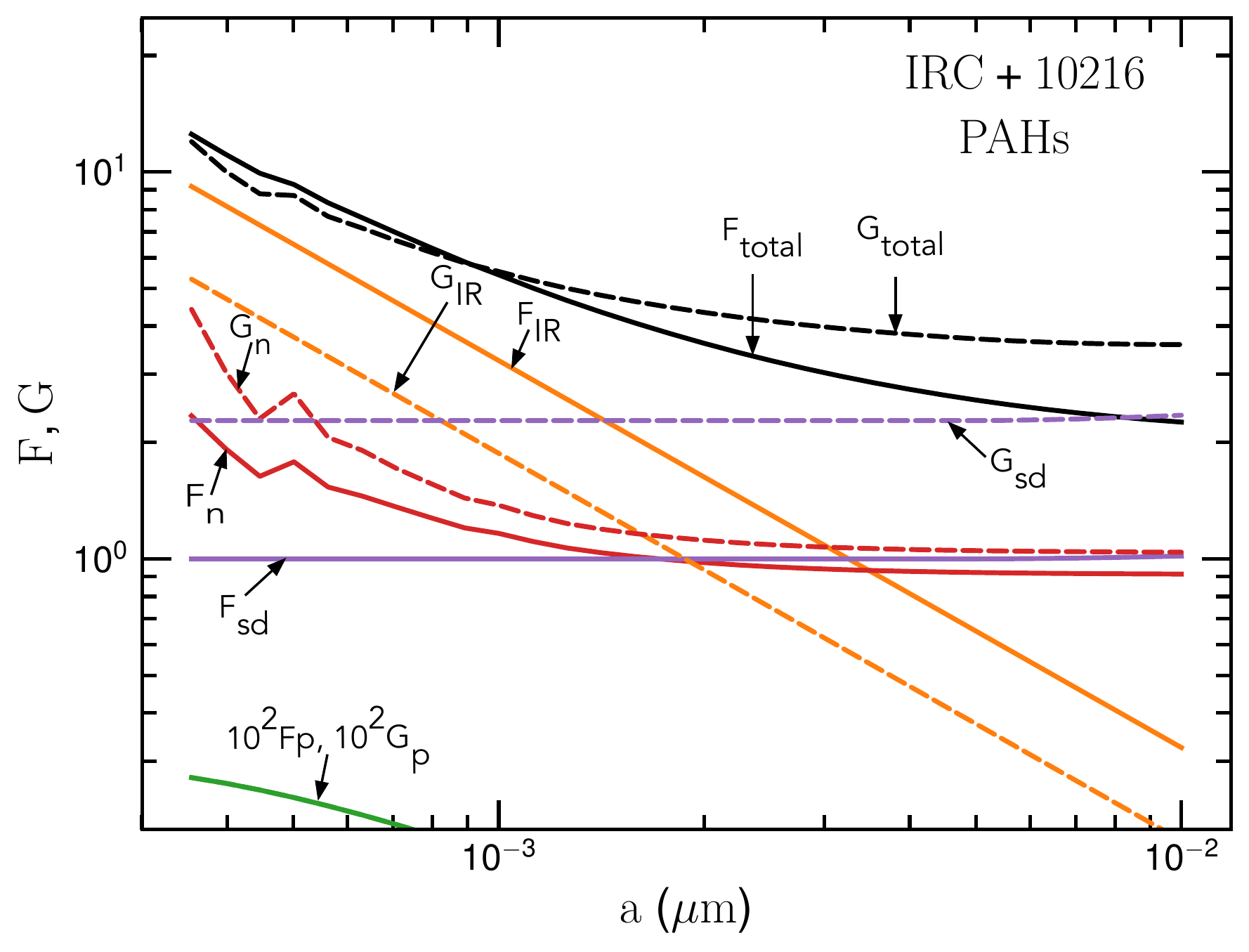}
\includegraphics[width=0.45\textwidth]{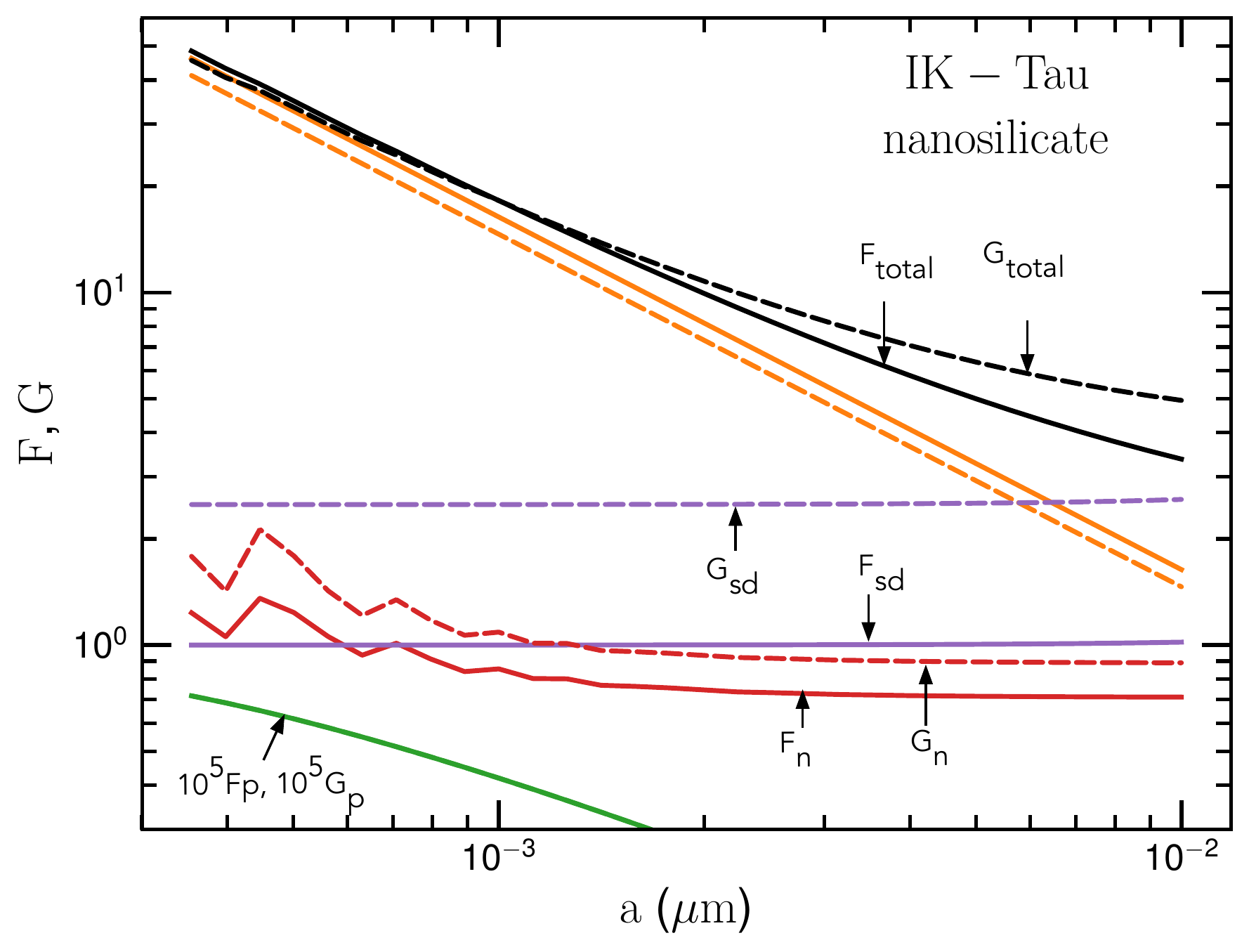}
\caption{Rotational damping ($F$, solid lines) and excitation ($G$, dashed lines) coefficients of various processes computed for PAH and nanosilicate at 10$R_{\ast}$.}
\label{fig:FG_a}
\end{figure}

Following \cite{1967ApJ...147..943J} and \cite{2019ApJ...877...36H}, one can define the dimensionless damping and excitation coefficients for interaction processes with respect to the damping and excitation coefficients of purely hydrogen neutral-grain collisions as:
\bea
\frac{d(I\omega_{r})}{dt} &=& \langle \Delta J'_{r}\rangle _{\H}\times \sum_{j}F_{j},\\
\frac{d (I\omega^{2})}{dt} &=& \frac{\langle (\Delta J')^{2}\rangle _{\H}}{I}\times G=\frac{3I\omega_{T}^{2}}{\tau_{\H}}\times \sum_{j}G_{j},\label{eq:Gcoeff}
\ena
where $\langle \Delta J'_{r}\rangle$ and $\langle (\Delta J')^2\rangle$ are the mean decrease of grain angular momentum along the r-axis and the mean increase of rotational energy per unit time by colliding with pure hydrogen, respectively, and $\omega_{T}$ is the thermal angular momentum of grains at temperature $T_{\rm gas}$. The dimensionless damping ($F$) and excitation ($G$) coefficients are defined respectively as:
\bea
&F =& F_{sd} + F_{IR} + F_{n} + F_{p} + F_{i} \\
&G =& G_{sd} + G_{IR} + G_{n} + G_{p} + G_{i}
\ena
where suffixes $sd$, $IR$, $n$, $p$, and $i$ stand for gas-grain drift, Infrared, neutral, plasma, and ion, respectively. $F_{sd}$ and $G_{sd}$ are computed as in \cite{1995ApJ...453..238R} (the details are described in Section 3 in \citealt{2019ApJ...877...36H}), whereas other coefficients are computed as described in Sections 4 and 5 in \cite{1998ApJ...508..157D}. Note that $F_{j}=G_{j}=1$ for grain collisions with purely atomic hydrogen gas. Figure \ref{fig:FG_a} shows the obtained values of $F$ and $G$ for various processes for PAH and silicates grains. As shown, IR emission is dominant for rotational damping of the smaller grains ($<0.005$ $\mu$m) and gas-grain drift is dominant for larger grains.  

Let $T_{\rm rot}$ be the rotational temperature of spinning nanoparticles, so that $I\langle\omega^{2}\rangle=3k_{B}T_{\rm rot}$. Thus, using the rms angular velocity from \cite{1998ApJ...508..157D}, we obtain
\bea
\frac{T_{\rm rot}}{T_{\rm gas}}=\frac{G}{F}\frac{2}{1 + [1+(G/F^2)(20\tH/3\ted)]^{1/2}},\label{eq:Trot}
\ena
where $\tH$ and $\ted$ are the characteristic damping times due to gas collisions and electric dipole emission (see \citealt{1998ApJ...508..157D}). As $\tau_{\ed}/\tau_{\H}\sim (a/3.5\AA)^{7}(n_{\H}/10^{4}\cm^{-3})$ (see \citealt{Hoang:2010jy}), this ratio is much bigger than 1 as long as $n_{\H}>10^{5}\cm^{-3}$. Therefore, $T_{\rm rot}/T_{\rm gas}\sim G/F$, i.e., the rotational temperature is only determined by the rotational damping ($F$) and excitation ($G$) coefficients.

Accordingly, the rotation rate at the rotational temperature $T_{\rm rot}$ is given by
\bea \label{eq:omega_Trot}
\frac{\omega_{\rm rot}}{2\pi}&=&\frac{1}{2\pi}\left(\frac{3k_{B}T_{\rm rot}}{I}\right)^{1/2}\nonumber\\
&\simeq& 1.4\times 10^{10}\hat{\rho}^{-1/2}a_{-7}^{-5/2}\left(\frac{T_{\rm rot}}{10^{3}\K}\right)^{1/2}\Hz.
\ena

\subsection{Rotational disruption of nanoparticles by stochastic mechanical torques}

To calculate the smallest size $a_{\rm min}$ that nanoparticles can withstand the rotational disruption, we compute $\langle \omega^{2}\rangle$ using the rotational temperature $T_{\rm rot}$ as given by Equation (\ref{eq:Trot}) at each radius for a grid of grain sizes from $0.35-10$ nm and compare it with $\omega_{\rm cri}$.

\begin{figure}
\includegraphics[width=0.45\textwidth]{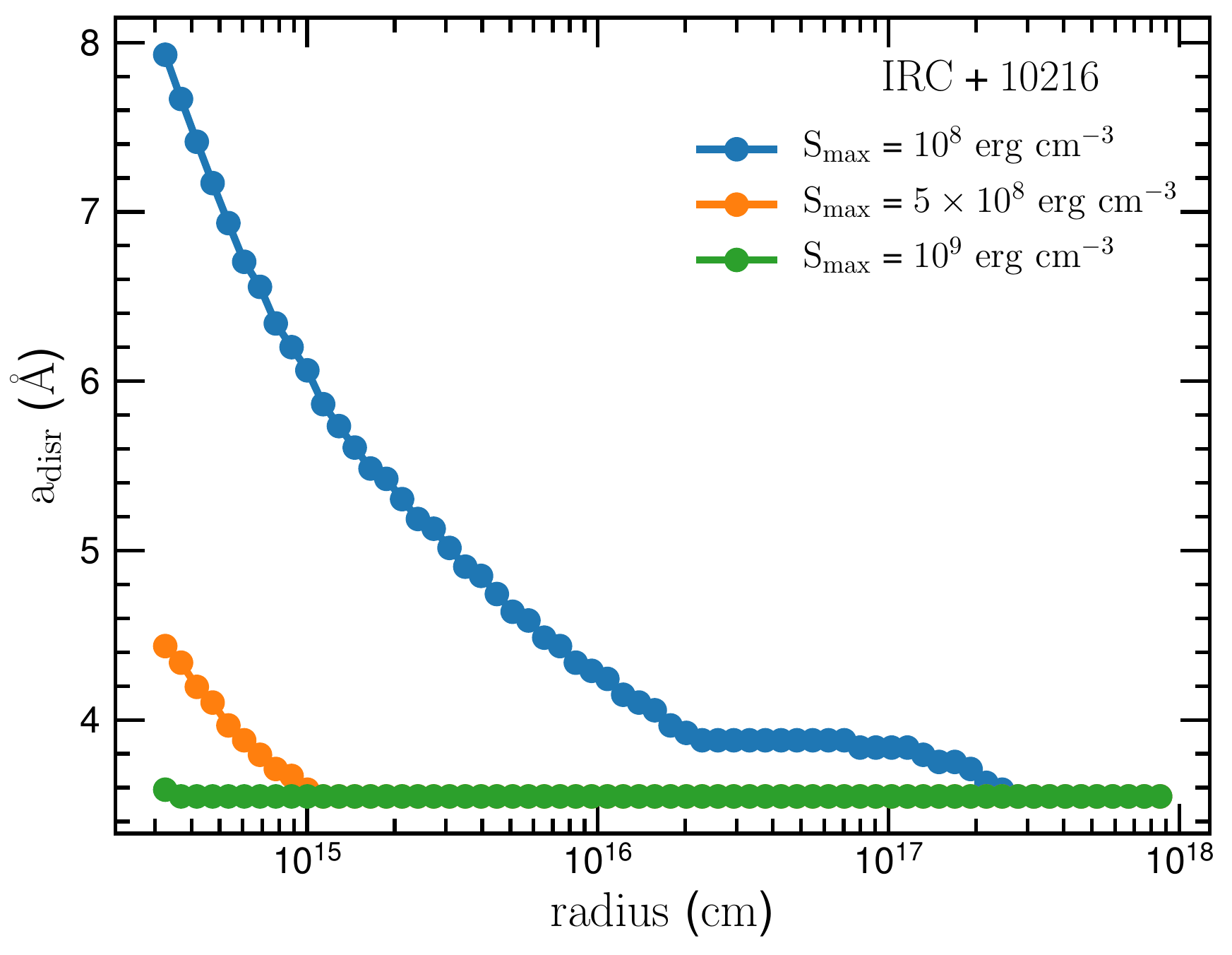}
\includegraphics[width=0.45\textwidth]{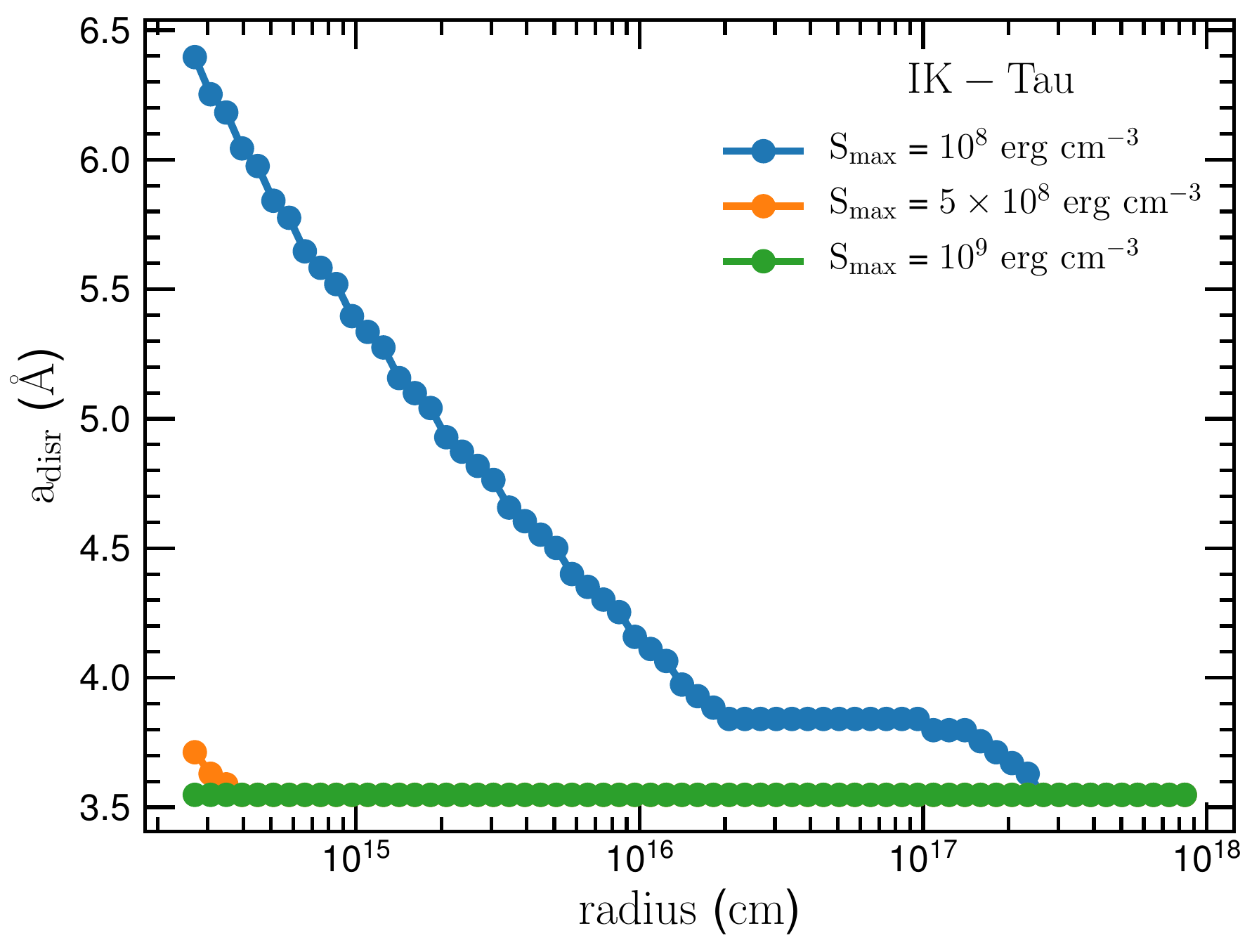}
\caption{Minimum size below which PAHs are destroyed by rotational disruption vs. radius in the CSE around IRC +10216 (top panel) and IK Tau (bottom panel), assuming the different material tensile strengths.}
\label{fig:a_cri}
\end{figure}

Figure \ref{fig:a_cri} shows the obtained minimum size $a_{\rm min}$ as a function of radius in CSE for different values of $S_{\max}$ and two AGB models. Strong nanoparticles can survive (green line), while weak nanoparticles can be destroyed (blue and orange lines). The disruption size drops gradually with distance from central star due to the rapid decrease of gas temperature with distance.

\section{Spinning dust emission}\label{sec:spin_model}
The emission from spinning dust is detailed in \cite{2019ApJ...877...36H}. In this section, we briefly recall the principles of this model.

\subsection{Size distribution of nanoparticles modified by rotational disruption} \label{ap:size_distr}
In the case without rotational disruption, we assume that dust grains in AGB envelopes consist of two populations, very small grains (nanoparticles) and larger grains, which we refer to as the {\it original dust populations}. In the presence of rotational disruption, smallest nanoparticles are removed by mechanical torques, whereas large grains are disrupted into smaller fragments which comprise nanoparticles, corresponding to the {\it modified dust population}. The size distribution of these modified dust populations is unknown, and below we adopt a simplified strategy to model their size distributions.

Since disruption by mechanical torques occurs at the lower end of the size distribution and disruption by radiative torques occur at the high end, we can account for these disruption effects separately. As the size distribution of the nanoparticles are poorly known, there is no strong conclusion whether this distribution should be a power-law or a log-normal. The power-law size distribution for nanoparticles was explored by \cite{2017ApJ...836..179H}. In this work, we assume the size distribution of {\it original} nanoparticles to be a log-normal as other studieds (e.g., \citealt{1998ApJ...508..157D}; \citealt{Li:2001p4761}; \citealt{2016ApJ...824...18H}):
\bea
    \frac{1}{n_\H}\frac{dn_{log-norm}}{da} = \frac{B}{a}\exp{\left( -0.5 \left[ \frac{log(a/a_0)}{\sigma}\right]^{2}\right)},
\ena
with $a_0=3-6\AA$, $\sigma=0.3-0.6$ the model parameters (see Tabel 1 in \citealt{Li:2001p4761}), and $B$ constant determined by $a_{0}$, $\sigma$ and $Y_{X}$ (fraction of total silicate ($X=Si$) or carbon ($X=C$) abundance contained in grains, see Eq 51 in \citealt{2019ApJ...877...36H}). The effect of rotational disruption by mechanical torques is then to increase the lower cutoff of the log-normal size distribution from $a_{\rm min}=3.5\AA$ to $a_{\rm cri}$ (see Figure \ref{fig:a_cri}).

To model the effect of rotational disruption by RATs, we assume that both {\it original} large grains and nanoparticles produced by disruption follow a power-law distribution with slope $\eta$:
\bea
    \frac{1}{n_\H}\frac{dn_{powerlaw}}{da} = A a^{\eta},\label{eq:dnda_power}
\ena
where $\rm A$ is a normalization constant determined by the dust-to-gas mass ratio ($M_{d/g}$) as:
\bea \label{eq:A_mody}
    A = \frac{(4+\eta)M_{d/g} m_{\rm gas}}{\frac{4}{3}\pi \rho_{\rm bulk} (a^{4+\eta}_{\rm max} - a^{4+\eta}_{\rm min}) }
\ena
with $a_{\rm max}$ the upper constraint  of grain size (Fig. \ref{fig:RATD}), and $a_{\rm min}$ the lower cutoff of grain sizes (Fig. \ref{fig:a_cri}). Since Equation \ref{eq:A_mody} is invalid for $\eta=-4$, the alternative expression of the constant A at this singularity is
\bea
    \lim_{\eta \to -4} A = \frac{M_{d/g}m_{gas}}{\frac{4}{3}\pi \rho_{bulk}(\ln a_{\rm max} - \ln a_{\rm min})}
\ena

For example, using a typical value of $M_{d/g}=0.01$, $\eta=-3.5$, $a_{\rm min}=50\AA$, and $a_{\rm max}=0.25\mu$m we estimate $A \simeq 10^{-25.16}\cm^{-2.5}$ for carbonaceous grains (as known as the MRN distribution, \citealt{1977ApJ...217..425M}). With these parameters, the contribution of the power-law distribution is negligible compared to the log-normal size distribution (see e.g., \citealt{1998ApJ...508..157D}). In the presence of RATD, large grains are disrupted into smaller ones, so we can extend the power-law to $a_{\rm min}=3.5\AA$. Hence, the size distribution (Eq. \ref{eq:dnda_power}) is modified such that the slope $\eta$ becomes steeper, and the contribution of the power-law becomes more important. Therefore, the net size distribution of nanoparticles is given by
\bea
\frac{dn}{da}=\frac{dn_{log-norm}}{da} + \frac{dn_{powerlaw}}{da}.\label{eq:dnda_total}
\ena
Since the slope $\eta$ is unknown, in the following, we will consider several values for $\eta$. The dust-to-gas mass ratio is likely to vary in practice, e.g., for IRC +10216: \cite{2006ApJ...650..374A} adopted $M_{d/g} = 0.002$, while \cite{2015A&A...575A..91C} used $M_{d/g} = 0.004$; or for IK Tau: \cite{2010A&A...516A..69D} adopted $M_{d/g} = 0.02$, while \cite{2016A&A...588A...4L} used $M_{d/g} = 0.01$.

\subsection{Spinning dust emissivity}
At any location in the CSE, the emissivity $j_{\nu}^{a}(\mu, T_{\rm rot})$ from a spinning nanoparticle of size $a$ is:
\bea
j_{\nu}^{a}(\mu, T_{\rm rot})=\frac{1}{4\pi}P(\omega,\mu)2\pi f_{\rm MW}(\omega),\label{eq:jem_a}
\ena
where $P(\omega,\mu)$ is the power emitted by a rotating dipole moment $\mu$ at angular velocity $\omega$ given by the Larmor formula,
\bea
P(\omega,\mu)=\frac{2}{3}\frac{\omega^4\mu^2\sin^2\theta}{c^3}~~~,
\ena
with $\theta$ the angle between spin vector $\bomega$ and moment vector
$\bmu$. Assuming a uniform distribution of $\theta$, then 
$\langle \sin^{2}\theta\rangle=2/3$. The dipole moment $\mu^{2}\simeq 86.5(\beta/0.4\D)^{2} a_{-7}^{3} \D^{2}$ for PAHs, and $\mu^{2}\simeq 66.8(\beta/0.4\D)^{2} a_{-7}^{3} \D^{2}$ for nanosilicates, in which $\beta$ is the dipole per atom \citep{2016ApJ...824...18H}. 
For PAHs, $\beta \simeq 0.4\D$, while nanosilicates are expected to have a large dipole moment depending on selection of molecules (see Table 1 in \citealt{2016ApJ...824...18H}). Therefore, we set $\beta$ as a free parameter when calculating spinning emission for nanosilicates grains. The grain angular velocity  $f_{\rm MW}$ can be appropriately described by a Maxwellian distribution in high-density conditions where collisional excitations dominate rotation of nanoparticles:
\bea
f_{\rm MW}(\omega, T_{\rm rot})=\frac{4\pi}{ (2\pi)^{3/2}}\frac{I^{3/2}\omega^{2}}{(k_{B}T_{\rm rot})^{3/2}}
\exp\left(-\frac{I\omega^{2}}{2k_{B}T_{\rm rot}} \right). \label{eq:fomega}
\ena

The rotational emissivity per H nucleon is obtained by integrating over the grain size distribution (see \citealt{2011ApJ...741...87H}):
\bea
\frac{j_{\nu}(\mu, T_{\rm rot})}{n_{\H}}=\int_{a_{\min}}^{100 \AA}j_{\nu}^{a}(\mu,T_{\rm rot})\frac{1}{n_{\H}} \frac{dn}{da} da,\label{eq:jem}
\ena 
where $dn/da$ is the size-distribution for PAHs and nanosilicates (see Section \ref{ap:size_distr}).

\subsection{Emission spectrum of spinning dust}
Assuming CSEs are spherically symmetric (see Figure 1 in \citealt{2010ApJ...711L..53S}), which is justified in the intermediate and outer region, i.e., $r>5R_{\ast}$ (e.g., , \citealt{2006ApJ...650..374A}; \citealt{2010A&A...516A..69D}), the
spectral flux density of spinning dust is: 
\bea
F_{\nu}= \frac{1}{4\pi D^{2}} \int_{R_{\rm in}}^{R_{\rm out}}dr 4\pi r^{2} n_{\H}(r)\left[\frac{4\pi j_{\nu}(\mu,T_{\rm rot})}{n_{\H}}\right],\label{eq:Fsd}
\ena
where $D$ is the distance from the AGB star to the observer (see Table \ref{tab:wind_params}).

\begin{figure}
\includegraphics[width=0.45\textwidth]{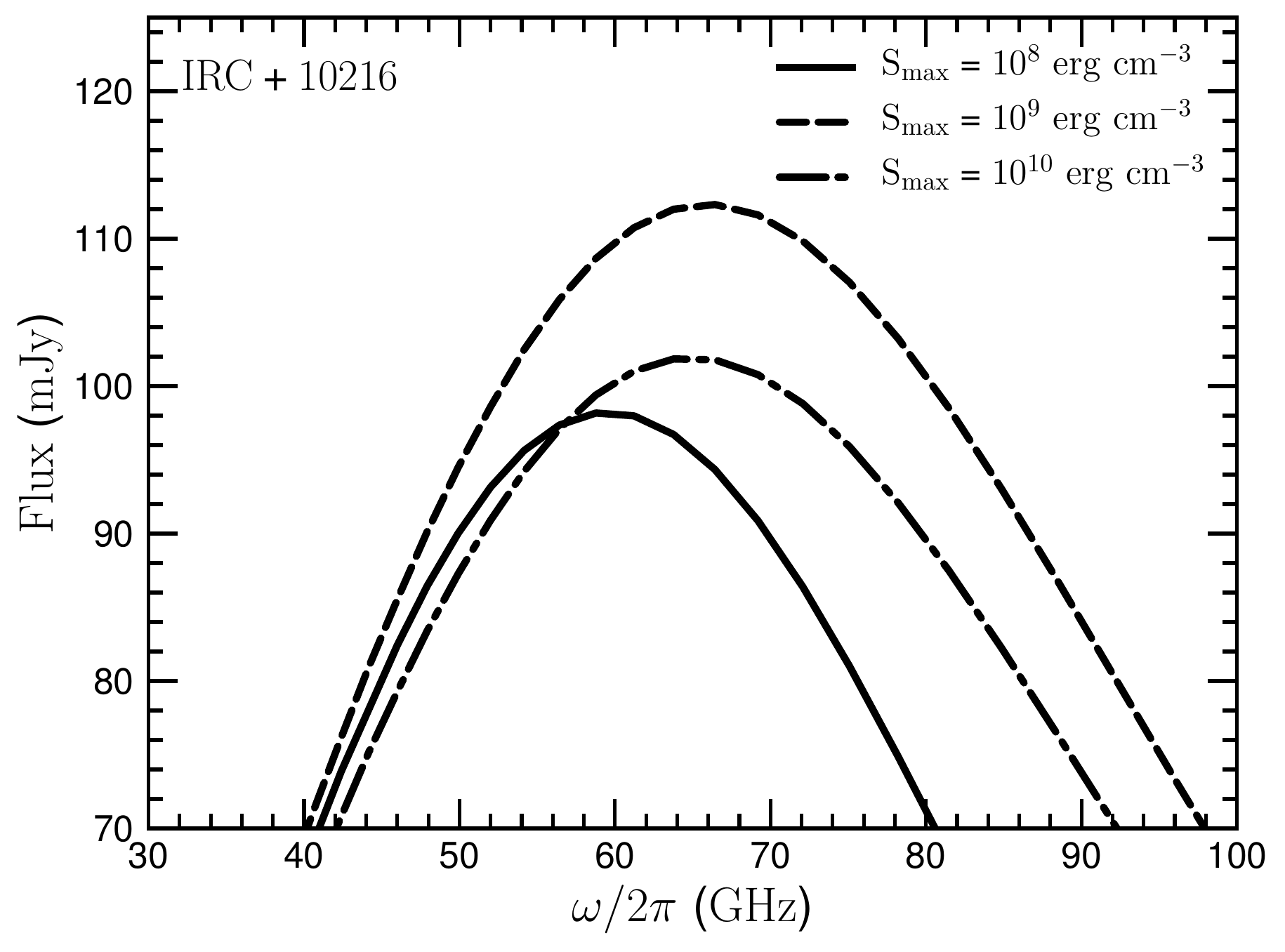}
\includegraphics[width=0.45\textwidth]{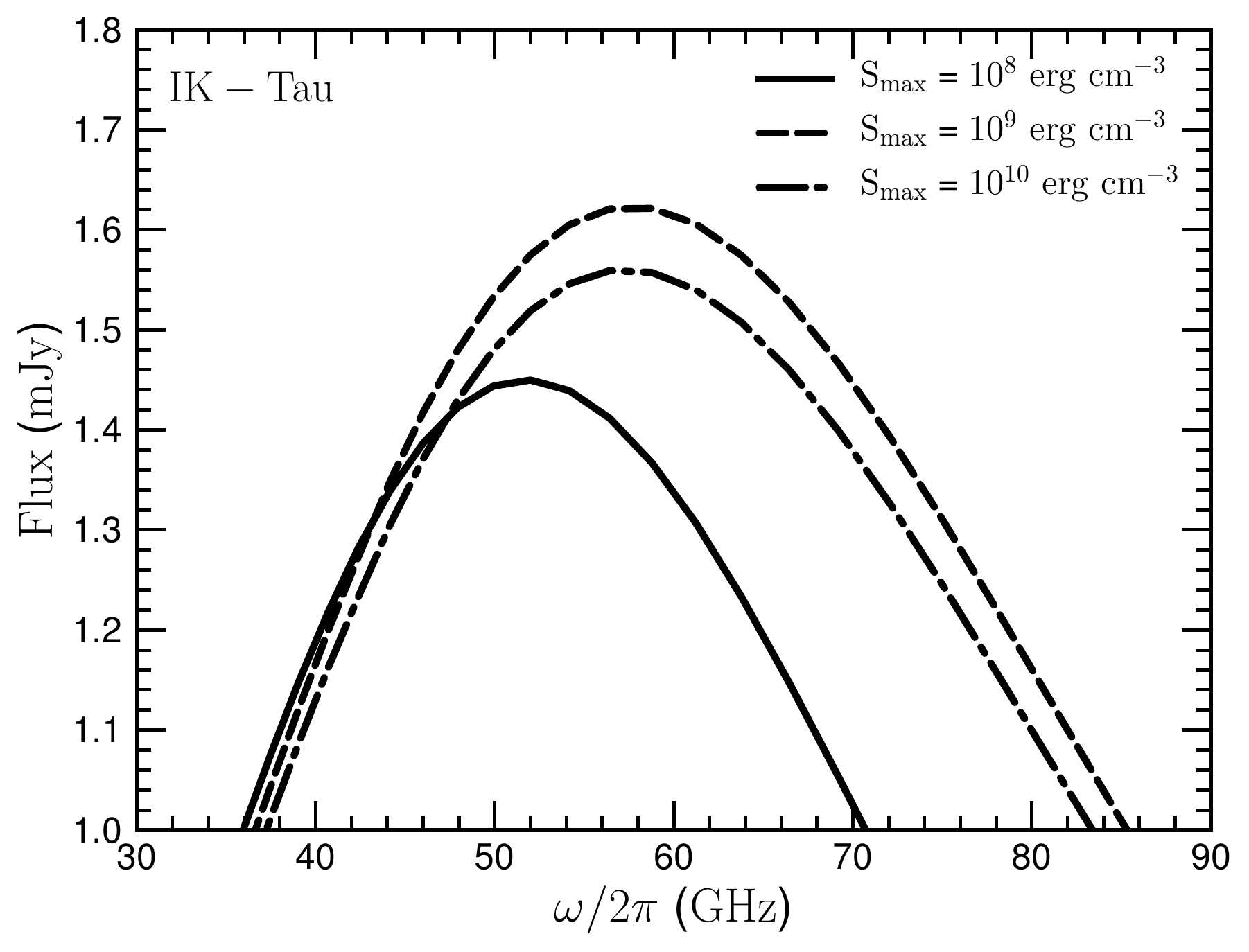}
\caption{Emission flux from spinning nanoparticles in CSEs around C-rich star (IRC+10216, top) and O-rich star (IK Tau, bottom). We adopt $M_{d/g}=0.01$, $\eta=-3.5$, $\beta=0.4\D$, $Y_C \sim 0.05$ (\citealt{2007ApJ...657..810D}), and $Y_{Si}=0.2$ (\citealt{2016ApJ...824...18H}).}
\label{fig:Flux}
\end{figure}

Figure \ref{fig:Flux} shows the emission spectrum of spinning PAHs in a C-rich star IRC +10216 (top panel) and nanosilicates in O-rich star IK Tau (bottom panel) for different values of material strengths. 
For strong grains (e.g., $S_{\rm max}=10^{10}\erg\cm^{-3}$) for which rotational disruption does not occur, the spinning dust emissivity is strong and can peak at high frequencies (see dashed lines). For weaker grains with rotational disruption, both the peak emissivity and peak frequency are reduced significantly because the smallest (and fastest-spinning) nanoparticles are destroyed into molecule clusters (see dashed-dotted and solid lines). 

\begin{figure}
\includegraphics[width=0.45\textwidth]{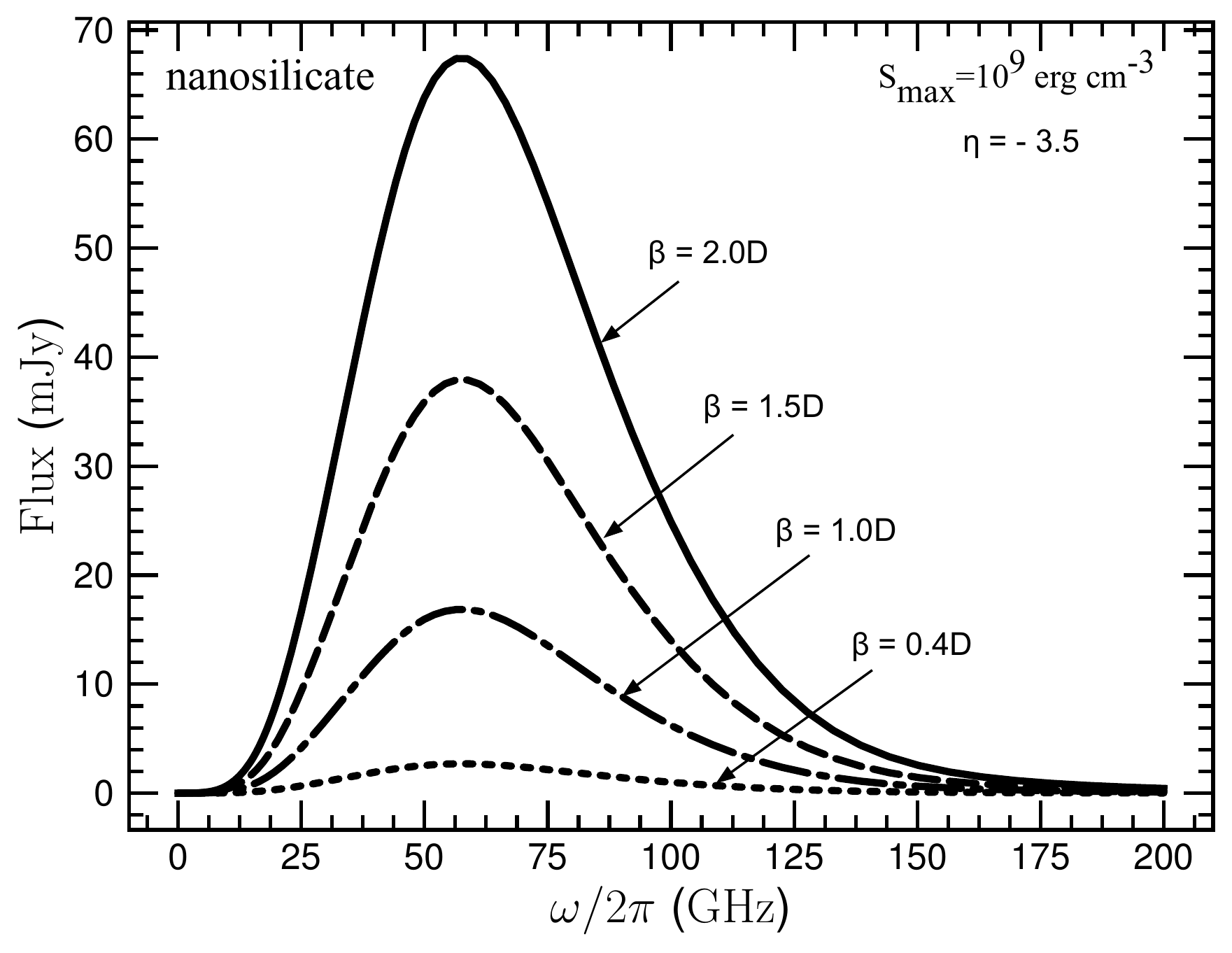}
\includegraphics[width=0.45\textwidth]{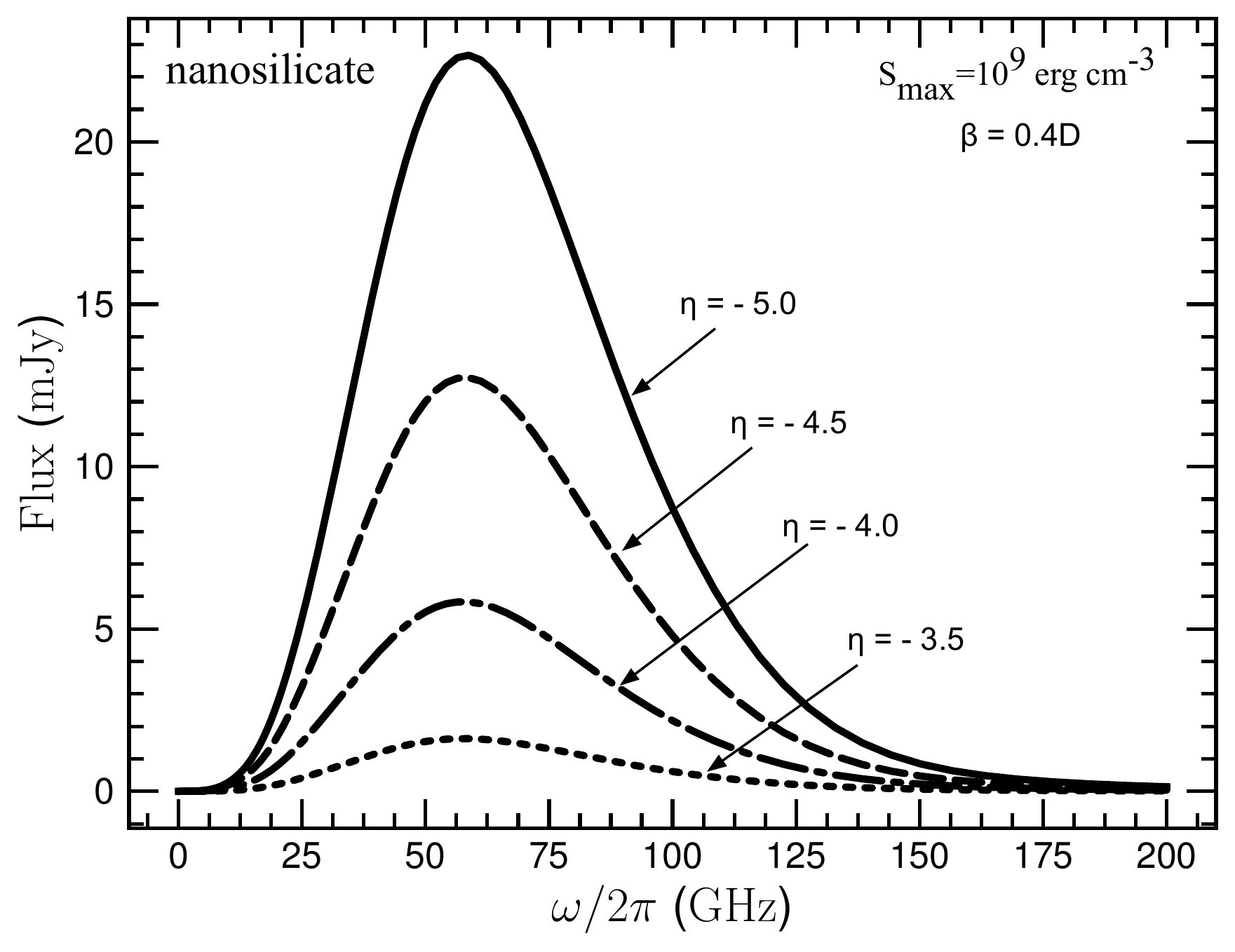}
\caption{Modification of the emission spectrum from spinning nanosilicates around IK Tau: (Top panel) varying the dipole per atom ($\beta$);
(Bottom panel) varying the slope of the size distribution ($\eta$). Here the tensile strength $S_{\max}=10^{9}\erg\cm^{-3}$ and $M_{d/g}=0.01$ are assumed.}
\label{fig:Flux_silicates}
\end{figure}

The spectrum of spinning dust emission depends on both the dipole moment and the size distribution.
The top panel of Figure \ref{fig:Flux_silicates} shows the increase of the emission flux from spinning nanosilicate grains with increasing the dipole moment because the power emitted by a rotating dipole moment is proportional to $\beta^{2}$. The peak frequency is insensitive to the $\beta$ value because in the dense envelopes, electric dipole damping is subdominant compared to gas damping.

The bottom panel of Figure \ref{fig:Flux_silicates} shows the dependence of spectrum on the slope of size-distribution, $\eta$. One can see that the peak emission flux tends to increase with increasing the magnitude of slope, but the peak frequency does not change.
Note that for the dust in the standard ISM, $\eta=-3.5$ (\citealt{1977ApJ...217..425M}). For circumstellar dust, \cite{1989A&A...223..227D} estimated this index to be $\sim -5$. The steeper size distribution might be explained by the enhancement of small particles owing to the disruption of large grains due to RATD mechanism. Therefore the constant $A$ of the grain-size distribution  increases for steeper slope (see Eq. \ref{eq:A_mody}) and results in an increment of flux.      

\begin{figure}
\includegraphics[width=0.46\textwidth]{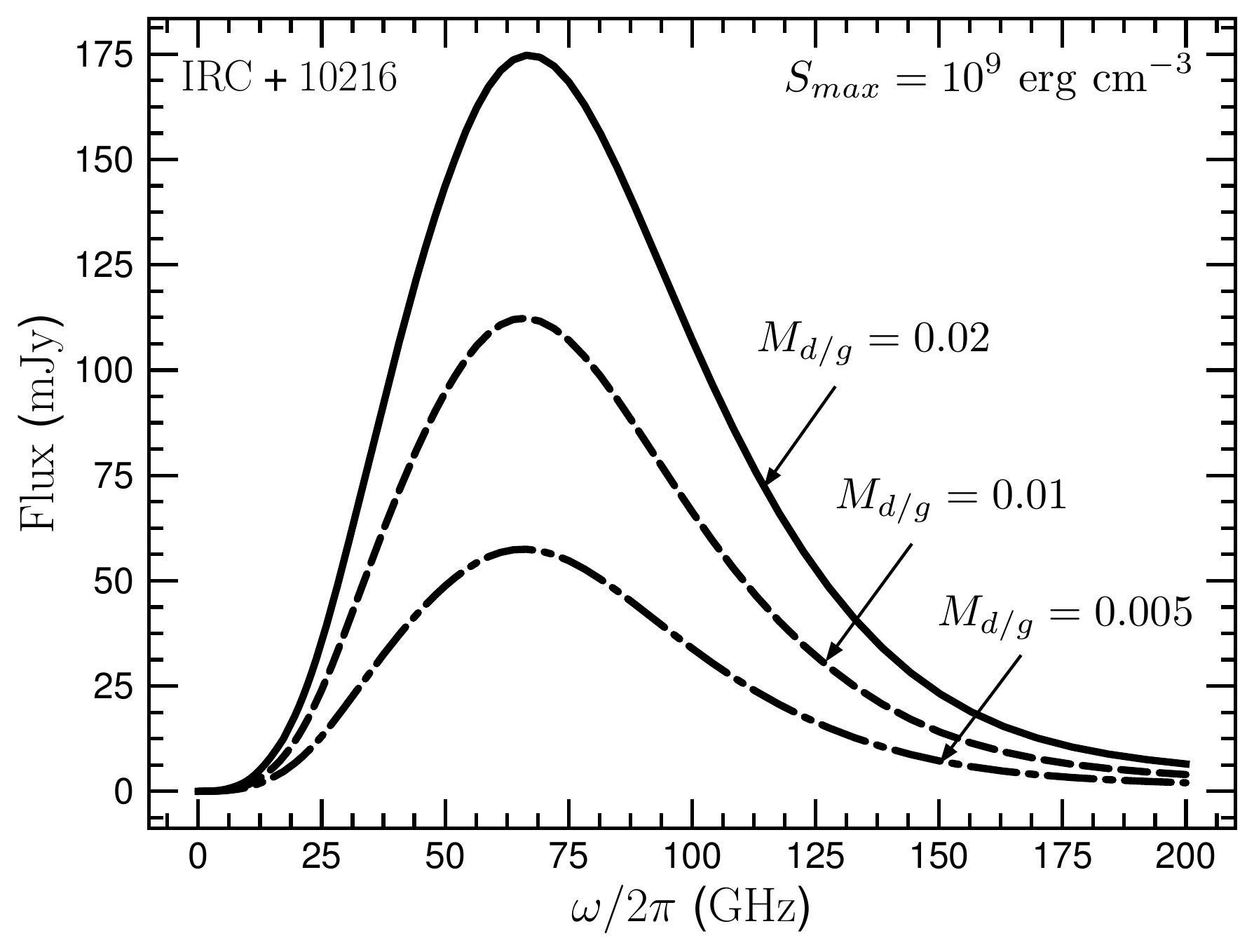}
\caption{Effect of dust-to-gas mass ratio on the emission spectrum of spinning grains. We adopt $\beta=0.4D$, $\eta=-3.5$}.
\label{fig:Flux_dust_to_gas}
\end{figure}

The dust-to-gas mass ratio ($M_{d/g}$) also affects on the spectrum of spinning dust emission. Figure \ref{fig:Flux_dust_to_gas} shows an example of the variation of spinning spectrum over some values of $M_{d/g}$ calculated in the IRC +10216. The higher value of the $M_{d/g}$ results the increment of the flux since the constant $A$ of the grain-size distribution is proportional to this quantity (see Eq. \ref{eq:A_mody}). 

\begin{figure}
\includegraphics[width=0.46\textwidth]{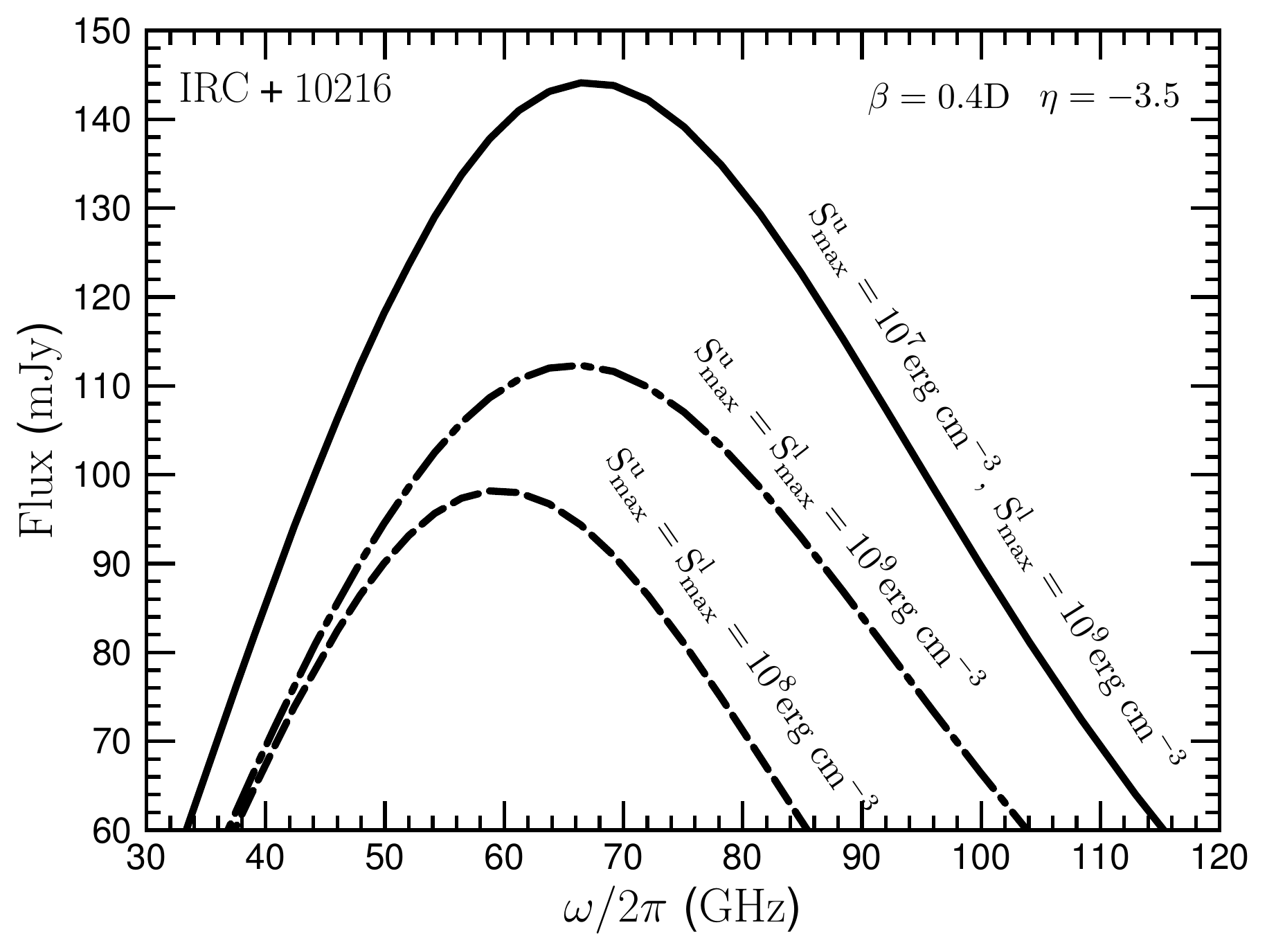}
\caption{Effect of size-dependent tensile strength on the emission spectrum of spinning grains. We adopt $\beta=0.4D$, $\eta=-3.5$, and $M_{d/g}=0.01$}.
\label{fig:Flux_Sl7Sm9}
\end{figure}

Above, we calculated the emission spectrum of spinning dust by using the same value of $S_{\max}$ for both large grains and small grains. In reality, the structure of small grains, however, would be more compact and that should cause a higher value of $S_{\max}$ than for the large ones. Figure \ref{fig:Flux_Sl7Sm9} shows the effect of the size-dependent tensile strength on the emission spectrum of spinning grains, in which we assume $S^{u}_{\max}=10^{7}\erg\cm^{-3}$ for the upper constraint on grains sizes by RATD, and $S^{l}_{\max}=10^{9}\erg\cm^{-3}$ for the lower constraint on those by mechanical torques. In this case, the emission flux has higher amplitude (solid line) than the cases of constant $S_{\max}=10^{9}\erg\cm^{-3}$. The underlying reason is owing to the enhancement of nanoparticles because weaker large grains (i.e., low value of $S_{\max}$) are more easy to be disrupted by RATD. Moreover, comparing to the case of the same $S^{l}_{\max}$ (dashed dotted line), one sees that the peak frequency does not change because the disruption effect of smallest nanoparticles by mechanical torques are the same in both cases. So, the peak frequency is much higher when comparing to the case of lower $S^{l}_{\max}$ (dashed line) as shown in Figure \ref{fig:Flux}.        

\section{Discussion}\label{sec:discuss}
\subsection{Rotational disruption of large grains into nanoparticles by radiative torques}
We have studied the rotational disruption of large grains spun-up by radiative torques using the theory developed by \cite{2019NatAs...3..766H}. As shown in Figure \ref{fig:RATD}, owing to their extremely fast rotation, large grains are being disrupted into smaller fragments, including nanoparticles, which places a constraint on the upper limit of the grain size distribution in CSEs around AGB stars. The efficiency of rotational disruption is different for different locations along the radial trajectory of stellar winds (see Sec. \ref{sec:RATD}). In addition, weak grains (i.e., tensile strength $S_{\max} \lesssim 10^{9}\erg\cm^{-3}$) are quite easily  disrupted, while the disruption of stronger grains (i.e., tensile strength $S_{\max} \gtrsim 10^{10}\erg\cm^{-3}$) is less efficient.       

\subsection{Removal of nanoparticles due to rotational disruption by mechanical torques}
Due to stochastic collisions with neutral and ionized gas, plasma drag, and infrared emission, nanoparticles tend to rotate thermally/subthermally ( \citealt{1998ApJ...508..157D}; \citealt{Hoang:2010jy}). Nevertheless, due to their small sizes (inertia moment), nanoparticles can rotate extremely fast, at rates more than $10^{10} \rm rad\s^{-1}$ (see Eq. \ref{eq:omega_Trot}). Subject to a supersonic gas flow induced by shocks (\citealt{2019ApJ...877...36H}; \citealt{2019ApJ...886...44T}) or radiation pressure as shown in this paper, nanoparticles can be spun-up to suprathermal rotation, resulting in the disruption of smallest nanoparticles into molecule clusters because the centrifugal stress exceeds the maximum tensile strength of nanoparticles. This mechanism places a contraint on the lower limit of the grain-size distribution. As shown in Figure \ref{fig:a_cri}, the disruption of nanoparticles is strongest near the central star and rapidly decreases outward. Moreover, nanoparticles of strong materials are hardly disrupted by mechanical torques in the AGB envelopes.   

\subsection{Can spinning dust explain excess microwave emission from AGB envelopes?}
 \begin{figure}
 \includegraphics[width=0.44\textwidth]{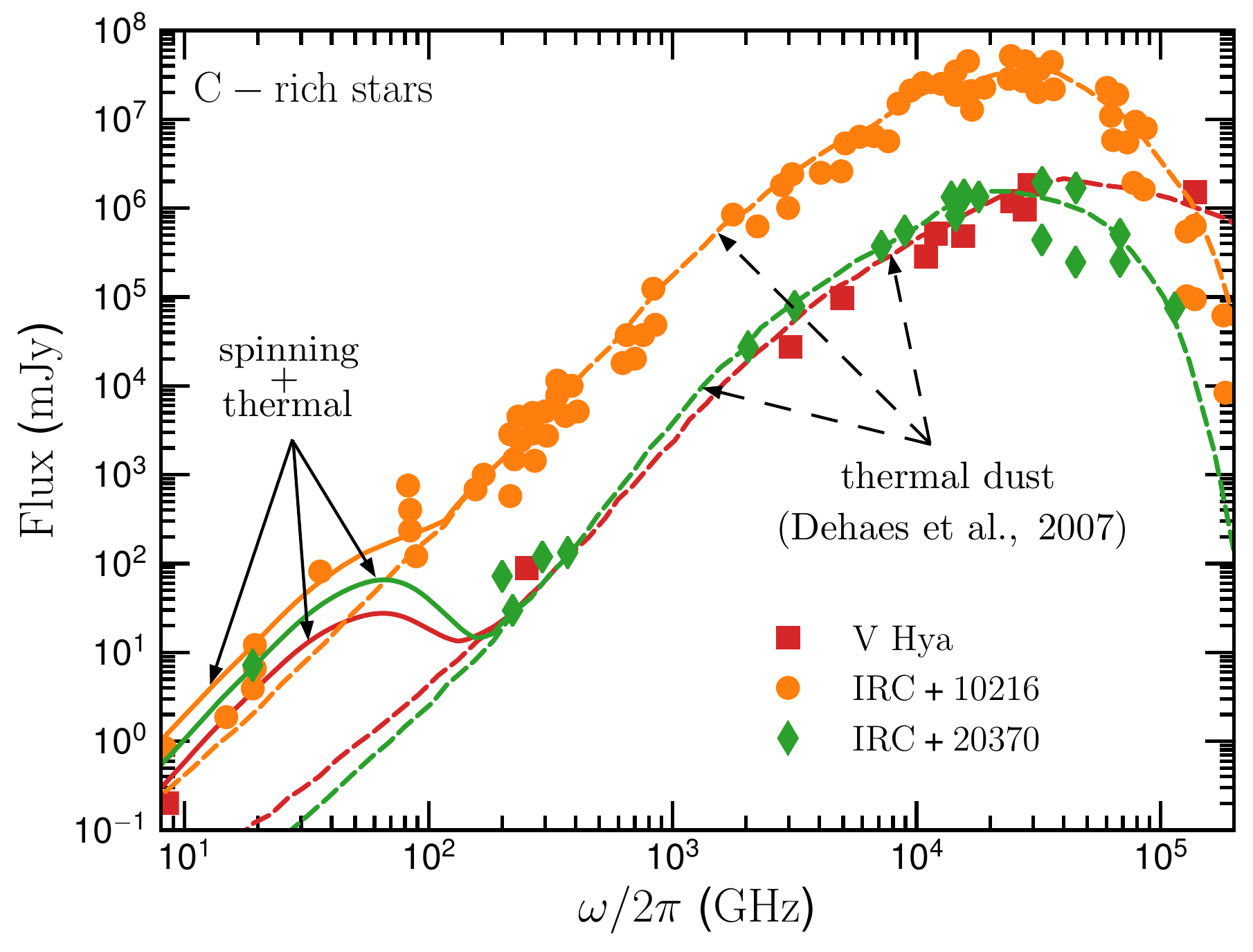}
 \includegraphics[width=0.45\textwidth]{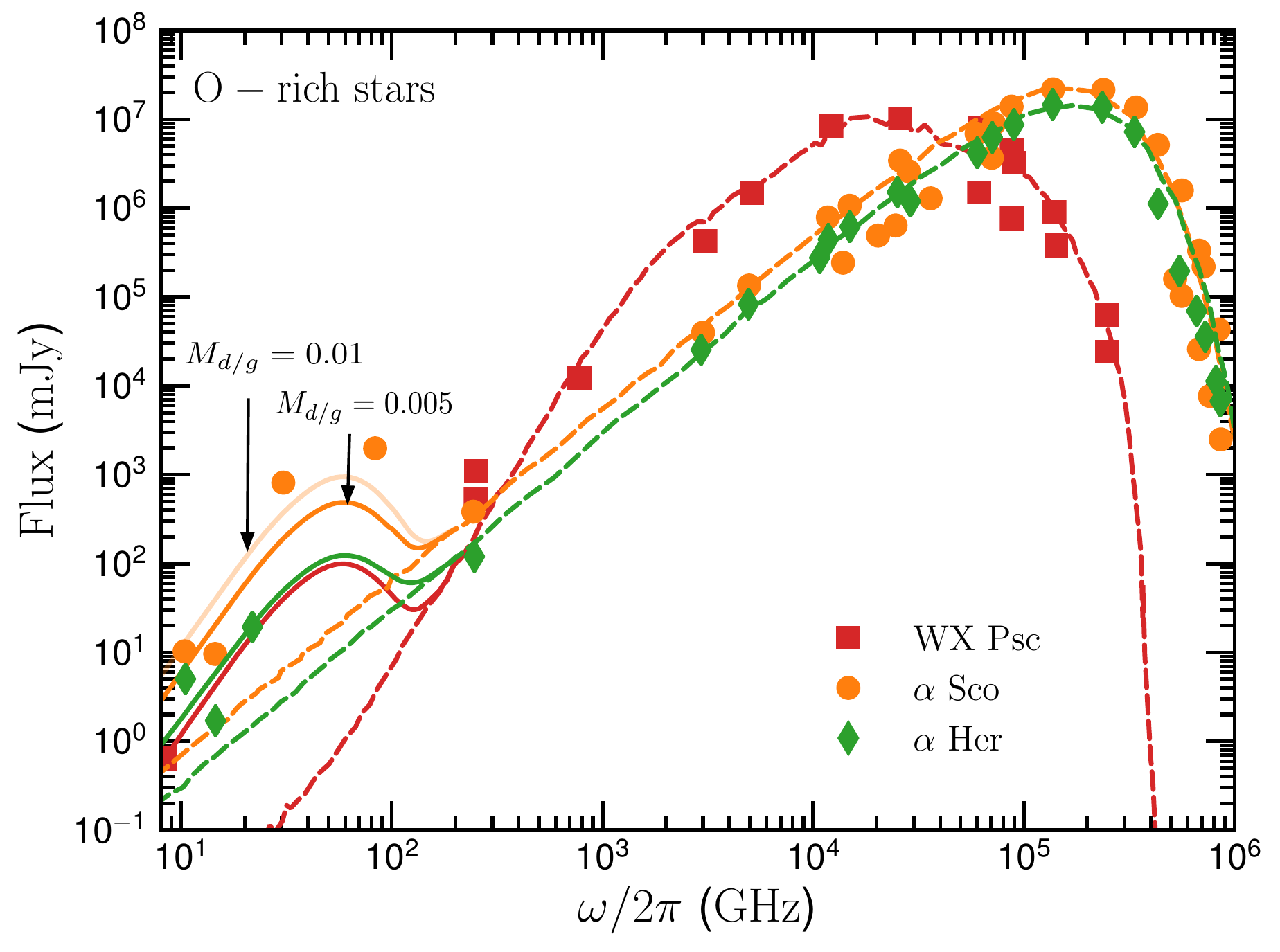}
 \caption{Comparison of spinning dust (solid lines, this work) and thermal dust (dashed lines, taken from \citealt{2007MNRAS.377..931D}) model to radio observational data from the literature (filled symbols). The solid lines show that the spinning dust model can reproduce well the CSEs of both C-rich (upper panel) and O-rich stars (lower panel). \textbf{C-rich stars} - $S_{\max} \geq 10^{10}\erg\cm^{-3}$, $\beta=0.4\D$: $\eta=-3.7$ (IRC +10216), $\eta=-3.6$ (IRC +20370) and $\eta=-3.2$ (V Hy$\alpha$). \textbf{O-rich stars} - $S_{\max} \geq 10^{9}\erg\cm^{-3}$: $\beta=2.3\D$, $\eta=-5.0$ ($\alpha$ Sco); $\beta=1.6\D, \eta \sim -4.4$ ($\alpha$ Her); $\beta=1.2 \D, \eta=-4.8$ (WX Psc). Note: the gas-to-gas mass ratio is fixed as $M_{d/g}=0.005$ as in \citealt{2007MNRAS.377..931D}. The faint orange line likely indicates a higher dust-to-gas-mass ratio of $M_{d/g}\simeq$0.01 in the case of $\alpha$ Sco.}
 \label{fig:AMEfit}
 \end{figure}
 
 The early detection of cm-wave observations toward to the AGB stars, i.e., at 15 GHz (or 2 cm) and 20 GHz (or 1.5 cm) from IRC+10216 (\citealt{1989A&A...220...92S}), and at 8.4 GHz (or 3.57 cm) from 4 AGB stars over 21 samples (\citealt{1995ApJ...455..293K}) cannot be explained by thermal dust emission. Recently, \cite{2007MNRAS.377..931D} presented the SED observations from a large sample of O-rich and C-rich AGB stars envelopes and showed emission excess at cm wavelengths for many stars, including some post-AGB and supergiants with circumstellar shells. The authors divided them into two groups. Group I could be fitted successfully with thermal dust. Group II, on the other hand, is being fitted very well at optical and IR bands, but not being reproduced at cm-mm bands, including: IRC +10216, $\alpha$ Her, IRC +20370, WX Psc, $\alpha$ SCo, and V Hya (with central peak at $\lesssim 100$ GHz); and AFGL 1922, IRAS 15194-5115 (with central peak at $\gtrsim 300 $ GHz).  

As shown in Section \ref{sec:spin_model}, the emission of spinning dust is strong and dominant over the thermal emission at the frequency $\lesssim 100$ GHz. It is worth to note that the global purpose of this work is to demonstrate the spinning dust is one possible explanation for the AME detections in AGB envelopes so that let us now use our model calculated for IRC +10216 and IK Tau to fit with the observational data of two C-rich and O-rich AGB stars, respectively. To fit the data, we vary three parameters $S_{max}$, $\beta$ and $\eta$ while fixing other physical parameters until we obtain the best-fit models. In purpose of showing the whole SED, we combine with the best model of the thermal dust emission provided by \cite{2007MNRAS.377..931D}, which were modelled by DUSTY code (\citealt{1999ascl.soft11001I}) with $M_{d/g}=0.005$. Figure \ref{fig:AMEfit} shows our best-fit models to observational data for three C-rich (upper panel) and three O-rich (lower panel) stars with given the corresponding set of dim parameters\footnote{The best fit parameters are reduced from the model of the spinning dust computed with the physical properties of IRC +10216 and IK Tau. Therefore, the exact parameters can be slightly different for other stars when we use their physical properties properly.} in the caption. Apparently, thermal dust and spinning dust are able to reproduce the mm-cm emission for both C-rich star (top panel, Figure \ref{fig:AMEfit}) and O-rich stars (bottom panel, Figure \ref{fig:AMEfit}). 

We note that some AGB stars such as AFGL 1922, IRAS 15194-5115 exhibit submm emission excess (i.e., at higher frequencies of $ \gtrsim 250$ GHz). Furthermore, \cite{2007MNRAS.377..931D} also reported the variability at 1.2mm ($\simeq$250GHz) for WX Psc and 1.1mm ($\simeq$ 273GHz) for IRC +10216, and suggested that the variable molecular line emission might explain these large relative variations. However, such submm excess cannot be reproduced by our standard spinning dust model presented here because spinning dust is known to be efficient at microwave frequencies. 
Very recently, \cite{2019ApJ...877...36H} and \cite{2019ApJ...886...44T} show that spinning dust can be efficient at $\nu\gtrsim 200$ GHz in magnetized shocks where nanoparticles can be spun-up to suprathermal rotation by supersonic neutral drift, provided that nanoparticles are strong enough to withstand disruption by centrifugal force.
Indeed,  termination shocks (or reversed shocks) could occur in CSEs when the stellar wind interact with the surrounding ISM because the terminal velocity of the wind is supersonic (about $\sim 10-30\rm km\ s^{-1}$). 
 Therefore, we propose that observing both the gas and the dust emission in the AGB envelopes should be crucial in order to have better interpretation to the observable submm emission excess because we could then have a better idea of where the termination shocks occur so that we could take the spinning dust emission from this shock into account. 

It is worth to mention that \cite{2011ApJ...739L...3S} suggested that the submm emission excess by means of thermal emission from cold, very large grains (above 1 mm) in post AGB or pre-PN. However, how dust can grow to such big sizes in AGB envelopes is difficult to reconcile, because theoretical calculations show that dust released by AGB outflows is small grains (see, e.g., \citealt{1994ApJ...434..713J}). 




\subsection{Implications for future observations}
Observations would help to distinguish the carrier of AME by comparing C-rich vs. O-rich stars, because PAHs are formed in C-rich AGB stars while silicates are formed in O-rich stars. The interferometers like ALMA and the VLA are capable of mapping the gas distribution around AGB stars and polarization observations to constrain the dust grain properties in the shell (\citealt{2018A&A...620A..75K}; \citealt{2019A&A...621A..50B}). Such investigations are few in number, but they could help understand dust nucleation and growth in wind-driven AGB stars. 

Figure \ref{fig:AMEfit} in this paper presents the spinning dust models fitted to available low-resolution radio observations of AGB stars. The low resolution, i.e., 1 data point (e.g., V Hya, WX Psc, IRC +20370) or few data points (e.g., a Her) makes the difficulty to distinguish whether the(se) excess point(s) resulted by the spinning dust or free-free or synchrotron or chromosphere emissions. High resolution dust continuum observations from ALMA in bands 3, 4, and 5 at $\sim$ 84, 125 and 163 GHz, respectively, can improve the constraints on fitted model parameter much better than done for available low-resolution observations, and the SED fitting can help resolve which mechanism (i.e., free-free, synchrotron, chromosphere, or spinning dust) is responsible for the AMEs in AGB envelopes. Additionally, carbon rich stars such as IRC +10216 can be observed for different transitions of CO molecules (e.g., \citealt{2015A&A...575A..91C}). The observations covering the ring structure will map the CO gas emission and hence the CSE of the star. Simultaneously, dust continuum observations at comparable resolution will be useful to map the dust shell and hence spatially correlating it with the gas emissions.
Multiband observations of AGB stars from VLA  with frequencies from 1-40 GHz may help in understanding the dust distribution by investigating the fluxes obtained in these frequencies and fitting SEDs. VLA plays an important role in covering the regimes of frequencies less than 40 GHz (see Figure \ref{fig:AMEfit}) at much higher resolution than that of available data in literature. Therefore, a combination of VLA and ALMA dust continuum observations in some of the bright AGB shells will help to pin down the dust characteristics using our dust models. 

VLA and ALMA also provide unique opportunities to map high resolution dust continuum and line polarization. We can investigate whether wind-swept shell region of these ABG stars is polarized. The polarization measurements in the envelope of AGB stars will help in investigation the dust-grain properties such as size and alignment efficiency. 

\subsection{Toward constraining internal structure of dust grains with microwave emission}
By modeling rotational disruption of grains and resulting microwave emission from spinning dust, we find that microwave emission has strong correlation with the tensile strength of grain materials. Both the peak flux and peak frequency tend to increase with increasing the tensile strength (see Figure \ref{fig:Flux_silicates}). This can be a powerful constraint on the internal structure of newly formed dust grains in AGB envelopes, which is still a mystery in dust astrophysics. Recently, \cite{2019ApJ...876...13H} suggested that the upper cutoff of the grain size distribution in the ISM can be constrained by rotational disruption.

\subsection{Model uncertainties}
In this section, we would like to stress the uncertainties and the limitations of our model to the readers. First, we neglect the optical depth effect; thus we overestimate the radiation field strength in the envelope that would result in the RATD effect on constraining the upper cutoff of grain size (see Fig. \ref{fig:RATD}) then change the grain size distribution (see Eq. \ref{eq:A_mody}). Second, we assume the dust temperature varies as in \cite{2003agbs.conf.....H} (see Eq. \ref{eq:Td}) deriving from the radiative equilibrium, instead of solving the radiative transfer. This assumption is not correct for the nanoparticles responsible for spinning dust since they should be transiently heated, and their temperature should be higher. However, as spinning dust emission depends mostly on rotation rate, which is different from thermal dust emission, the effect of higher $T_{d}$ is to increase thermal fluctuations of the principal grain axis with its angular momentum. The thermal fluctuation is ineffective at the high value of dust temperature (\citealt{2011ApJ...741...87H}). Third, we adopted the angular velocity follows the Maxwellian distribution to calculate the spinning dust emissivity as \cite{1998ApJ...508..157D} for the sake of simplicity. For more general, \cite{2009MNRAS.395.1055A} used the Fokker-Planck equation to derive this distribution, the predictions are not so much different from \cite{1998ApJ...508..157D} nevertheless. Forth, as \cite{1998ApJ...508..157D} and \cite{2009MNRAS.395.1055A}, our model disregarded the non-sphericity of grains and the anisotropy in the damping and excitation processes (i.e., ignore the magnetic field). Taking them into account, \cite{Hoang:2010jy} indicates that the peak emissivity and the peak frequency of the spinning dust emission increase by a factor of few. Fifth, since we do not know exactly the value of $Y_{Si}$ in AGB envelopes, we fix the value of the $Y_{Si} = 20\%$ as an arbitrary parameter in this work within a reference that \cite{Li:2001p4761} indicated the upper limits for $Y_{Si}$ in diffuse ISM for amorphous ultra-small silicate grains is 30$\%$. Moreover, the effect of $Y_{Si}$ has shown in Figure 2 in \cite{2016ApJ...824...18H}, which cause only the variation of the emission flux. These uncertainties can modify the best-fit parameters in Figure \ref{fig:AMEfit}, which urges the necessary to improve the model, and the specific model for each star is desired. 

Despite of these uncertainties, we believe that our model still shows the principle characteristics of spinning dust, and that the spinning dust can be a good mechanism to explain the AME emissions in AGB envelopes.

\section{Summary}\label{sec:sum}
We have studied rotational disruption of dust grains by radiative and mechanical torques in the AGB envelopes, performed detailed modeling of microwave emission from rapidly spinning nanoparticles, and applied the models to explain the observed excess microwave emission. The principal results are summarized as follows:

\begin{itemize}

\item[1]
We model the rotational disruption of large grains by centrifugal stress induced by radiative torques from central star. We find that large grains (e.g., $a>0.1\mum$) made of weak materials (tensile strength $S_{\max}\lesssim 10^{9}\erg\cm^{-3}$ can be disrupted into nanoparticles within a radius of $10^{16}\cm$ from the star.

\item[2] We also study the disruption of nanoparticles by centrifugal stress due to stochastic collisions of grains with supersonic gas flow driven by radiation pressure and find that smallest nanoparticles of weak materials located close to the star can be destroyed.

\item[3]

We model microwave emission from spinning PAHs and silicate nanoparticles in C-rich and O-rich envelopes. We find that due to the radial dependence of the gas temperature, the spinning dust emits over a wide range of microwave frequencies.

\item[4] 
We found that microwave emission from either spinning PAHs or spinning nanosilicates can dominate over thermal dust at frequencies $\nu<100$ GHz in the AGB envelopes.

\item[5] 
By fitting the spinning dust to observed data from mm-cm wavelengths, we find that AME observed in AGB envelopes can be successfully reproduced by microwave emission from carbonaceous or silicate nanoparticles.

\item[6] Thanks to the correlation of spinning dust flux with the grain tensile strength, we suggest that internal structure of newly formed dust in AGB envelopes can be probe with microwave emission observations.

\end{itemize}
\acknowledgments
We thank the anonymous referee for helpful comments that improve the impact and the presentation of this paper. This work was supported by NASA through award $\#$06 0001 issued by USRA, Basic Science Research Program through the National Research Foundation of Korea (NRF) funded by the Ministry of Education (2017R1D1A1B03035359), and National Science Foundation Grant-1715876. A.S acknowledges the financial support from the NSF through grant AST-1715876.  

\bibliographystyle{/Users/thiemhoang/Dropbox/Papers2/apj}
\bibliography{/Users/thiemhoang/Dropbox/Papers2/cites_paperApJ,/Users/thiemhoang/Dropbox/Papers2/cites_Books}

\end{document}